\documentclass[journal]{IEEEtran}
\usepackage{amsmath,amsfonts}
\usepackage{algorithmic}
\usepackage{algorithm}
\usepackage{array}
\usepackage[caption=false,font=normalsize,labelfont=sf,textfont=sf]{subfig}
\usepackage{textcomp}
\usepackage{stfloats}
\usepackage{url}
\usepackage{verbatim}
\usepackage{graphicx}
\usepackage{cite}
\usepackage[nolist]{acronym}
\usepackage{xcolor}
\usepackage{multirow}
\usepackage[normalem]{ulem}
\begin{document}
% \SetWatermarkText{Pre-Print}  % Replace with your watermark text
% \SetWatermarkScale{1}            % Scale of the watermark (adjust as needed)
% \SetWatermarkColor[gray]{0.8}      % Color of the watermark; 0.8 is light gray

% \title{Synergizing Channel-Based  QoT-Aware Resources Assignment for  Cross-Layer Planning in Multi-band IPoEONs}

\title{Synergizing Hyper-accelerated Power Optimization and Wavelength-Dependent QoT-Aware Cross-Layer Design in Next-Generation Multi-Band EONs}

% \author{Farhad Arpanaei,~\IEEEmembership{Member,~IEEE,}, Kimia Ghodsifar, Hamzeh Beyranvand, Mahdi Ranjbar Zefreh, Antonio Napoli, Luis Velasco, Jao Pedro, David Larrabeiti, Juan Pedro Fernández-Palacios

% \author{Farhad Arpanaei,~\IEEEmembership{Member,~IEEE,}, Kimia Ghodsifar, Hamzeh Beyranvand, Mahdi Ranjbar Zefreh, ...
%, Marija Furdek
\author{Farhad Arpanaei, Mahdi Ranjbar Zefreh, Yanchao  Jiang, Pierluigi  Poggiolini, Kimia Ghodsifar, Hamzeh Beyranvand, Carlos Natalino, Paolo Monti, Antonio Napoli, José M. Rivas-Moscoso,  \'{O}scar Gonz\'{a}lez de Dios, Juan P. Fernández-Palacios, Octavia A. Dobre, José Alberto Hernández, and David Larrabeiti
        % <-this % stops a space
\thanks{F. Arpanaei, J.A Hernández, and D. Larrabeiti are with the Department of Telematic Engineering, Universidad Carlos III de Madrid (UC3M), Leganés,
Madrid, 28911 Spain e-mail: farhad.arpanaei@uc3m.es.

M. Ranjbar Zefreh is with CISCO Systems S.R.L., Vimercate (MB), Italy.

Y. Jiang and P. Poggiolini are with Politecnico di Torino, Turin, Italy.

K. Ghodsifar and H. Beyranvand are with the Department of Electrical Engineering, Amirkabir University of Technology (Tehran Polytechnic), Tehran, Iran.

C. Natalino and  P. Monti are with the Electrical Engineering Department, Chalmers University of Technology, 41296 Gothenburg, Sweden.

A. Napoli is with Infinera, Munich, Germany.

J.M Rivas-Moscoso,  O. Gonz\'{a}lez de Dios, and J.P Fernández-Palacios are with Telefónica Global CTIO, S/N, 28050 Madrid, Spain.

Octavia A. Dobre is with Memorial University, Canada.

Farhad Arpanaei acknowledges support from the CONEX-Plus program funded by Universidad Carlos III de Madrid and the European Union's Horizon 2020 research and innovation program under the Marie Sklodowska-Curie grant agreement No. 801538. The authors from UC3M, Telefonica, and Infinera would like to acknowledge the support of the EU-funded ALLEGRO project (grant No.101016663). The authors from UC3M would
like to acknowledge the support of the Spanish-funded Fun4date-Redes project (grant No.PID2022-136684OB-C21).}

% \thanks{M. Ranjbar Zefreh is with CISCO Systems S.R.L., Vimercate (MB), Italy.}

% \thanks{Y. Jiang and P. Poggiolini are with Politecnico di Torino, Turin, Italy.}

% \thanks{K. Ghodsifar and H. Beyranvand are the Department of Electrical Engineering, Amirkabir University of Technology (Tehran Polytechnic), Tehran, Iran.}

% \thanks{C. Natalino and  P. Monti are with the Department of Electrical Engineering, Chalmers University of Technology, 41296 Gutenberg, Sweden.}

% \thanks{A. Napoli is with Infinera, Munich, Germany.}

% \thanks{J.M Rivas-Moscoso,  O. Gonz\'{a}lez de Dios, and J.P Fernández-Palacios are with Telefónica Research and Development, Ronda de la Comunicación, S/N, 28050 Madrid, Spain.}

% \thanks{ Octavia A. Dobre is with Memorial University, Canada.}
  
 \thanks{Manuscript received March 31, 2024; }
}

% Farhad Arpanaei, Mahdi Ranjbar Zefreh, Yanchao  Jiang, Pierluigi  Poggiolini, Kimia Ghodsifar, Hamzeh Beyranvand,  Carlos Natalino, Paolo Monti, Antonio Napoli, José M. Rivas-Moscoso,  \'{O}scar Gonz\'{a}lez de Dios, Juan Pedro Fernández-Palacios, José Alberto Hernández, and David Larrabeiti
% \thanks{F. Arpanaei, J.A Hernández, and D. Larrabeiti are with the Department of Telematic Engineering, Universidad Carlos III de Madrid (UC3M), Leganés,
% Madrid, 28911 Spain (e-mail: {farpanae, jahgutie, and dlarra }@it.uc3m.es), M. Ranjbar Zefreh is with CISCO, Milan, Italy (email: mranjbar@cisco.com),Y. Jiang and P. Poggiolini are with Politecnico di Torino, Turin, Italy (email: ....), K. Ghodsifar and H. Beyranvand are with .....   (email: ... ), C. Natalino and  P. Monti are  with .... (email: ... ), Gutenberg, Sweden, A. Napoli is with ....., Munich, Germany (email: ), J.M Rivas-Moscoso,  O. Gonz\'{a}lez de Dios, J.P Fernández-Palacios are with ...., Madrid, Spain (emails: ....)  .

\markboth{IEEE Journal on Selected Areas in Communication (Pre-Print)}%
{F\'Arpanaei\'an \MakeLowercase{\textit{et al.}}:Synergizing Hyperaccelerate Power Optimization}


\maketitle

\begin{abstract}
The extension of elastic optical network technologies
to multi-band transmission (i.e., MB-EON) holds significant
promise in augmenting spectral efficiency, total throughput, and
reducing the total cost of ownership over the long term for
telecommunications operators. Nevertheless, the design intricacies of such networks engender several challenges, at the forefront
of which is the optimization of physical parameters, e.g., optical
power and the corresponding quality of transmission (QoT). At
the heart of these challenges are frequency-dependent characteristics of the fiber, such as fiber loss, dispersion, nonlinear coefficients, and the interference effects of inter-channel stimulated
Raman scattering is particularly pertinent when operating beyond
the L+C (LC)-band when occupying a continuous transmission
spectrum wider than 100 nm. In this investigation, we introduce
a methodology to find the optimal optical power allocation on
a span-by-span basis. Two hyper-accelerated power optimization
(HPO) strategies have been introduced: flat launch power (FLP)
and flat received power (FRP). These methodologies offer a
significant acceleration in network power optimization compared
to existing approaches while maintaining the stability of running
services. Furthermore, we exhaustively compare the FLP and
FRP models. Our research reveals that while FRP does not
notably increase total capacity (showing an increase of less than
10 Tbps for an L+C+S (LCS)-band system spanning 100 km), its
the impact is negligible for systems operating in the C- and LC-band
scenarios. However, in the LCS scenario, FRP brings notable
improvements in flatness (the difference between maximum and
minimum values) and the minimum of the generalized signal-
to-noise ratio (GSNR)/optical SNR (OSNR), especially in the S-
band, achieving approximately 2/0 dB and 2.5/6 dB, respectively.
In a network-wide analysis of several network topologies, we demonstrate that the improvement in the minimum GSNR due to the FRP technique, synergizing with wavelength-dependent QoT-aware cross-layer design, results in a throughput increase ranging from approximately 12\% to 75\% (depending on the network scale) at a 1\% bandwidth blocking rate. Lastly, we apply HPO to local and global power optimization methods in MB-EON, showing that, while both methods exhibit similar performance, the latter is simpler and more cost-effective for larger-scale networks.
\end{abstract}

% \AntoComm{I would actually cite something like Agrawal.}
\section{Introduction}\label{sec_I_intro}
\IEEEPARstart{T}{he} multi-band (MB) technology is a cost-effective solution to address the growing demand for higher bandwidth in metro and core networks, which experience an annual traffic demand increase of more than 35\%-40\% \cite{Souza_cost24, Farhad_ICTON2023,Jana2023}. However, the adoption of this technology is currently in the early stages, with only the L+C (LC)-band having been recently commercialized \cite{hoshida2022ultrawideband,InfineraCL}. In addition to the hardware requirements, associated with implementing the LC-band and beyond, such as amplifiers and transceivers, the planning of MB optical networks require addressing the inter-channel stimulated Raman scattering (ISRS) effects. ISRS causes the power of high-frequency channels to deplete into the low-frequency channels. This phenomenon becomes particularly significant when the dense wavelength division multiplexing (DWDM) bandwidth is wider than 10 THz \cite{Semrau2019}. 
Network planning becomes even more complex in EONs when the choice of the modulation format for a lightpath varies based on the reach distance, number of spans, and the number of traversed reconfigurable optical add-drop multiplexer (ROADM) components.

The planning and service provisioning of MB-EONs has been studied over the past six years covering four main areas: 1- Migration scenarios from the C-band to beyond C-band, including C to LC/LCS/LCS+E/LCS+E+O/U+LCS+E+O transitions, e.g., \cite{Souza_cost24, SamboJLT2020,6DMANJOCN2024}, 2- Techno-economic studies, e.g., \cite{6DMANJOCN2024,Jana2023, Souza_cost24}, 3- Power optimization and control strategies, e.g., \cite{Semrau_PowerOptimization, Lasagni_ECOC2021, souza2022optimal,Souza_Genetic, Correia_PowerControl,khan_2017,Buglia2022,Luo_22_optic_express,RanjbarECOC22022}, and 4- Routing, modulation format, band, and spectrum/wavelength assignment algorithms (RMBS(W)A), e.g., \cite{MehrabiJLT2021,SamboJLT2020,Re_Verification2022JLT}. 
Given the focus of this paper on greenfield network planning for MB-EONs, we will not delve into the different migration strategies (e.g., day one and pay as you grow) or techno-economic studies.
Launch power is one of the most critical parameters influencing the generalized signal-to-noise ratio (GSNR) value, which serves as a QoT metric \cite{Correia_PowerControl}.
Regarding the power optimization, the previous works have considered various objective functions such as link capacity (i.e., achievable information rate) maximization such as \cite{Semrau_PowerOptimization, souza2022optimal, RanjbarECOC22022}, minimum GSNR maximization like \cite{Lasagni_ECOC2021}, or flatness of the GSNR profile in each band or across all bands such as \cite{Souza_Genetic, Correia_PowerControl, Poggiolini_OFC_2024}.
However, solving the power optimization problem while considering the ISRS effects is challenging, as it is NP-hard and non-convex \cite{Semrau_PowerOptimization, khan_2017}. 
To address this challenge, various brute-force heuristic and meta-heuristic algorithms have been proposed, including genetic algorithm \cite{Souza_Genetic}, particle swarm optimization \cite{Semrau_PowerOptimization,Buglia2022}, and the greedy search \cite{Luo_22_optic_express, Correia_PowerControl}. 
Additionally, while some authors have tackled the problem on a channel-by-channel basis \cite{khan_2017, Semrau_PowerOptimization}, others have adopted a span-based strategy by considering auxiliary parameters.
For instance, in \cite{Luo_22_optic_express} and \cite{Correia_PowerControl}, the authors proposed using two auxiliary parameters for each band: fixed tilt (slope) and offset, known as the (tilt, offset) approach. Additionally, in \cite{RanjbarECOC22022}, four parameters - offset, slope, parabolic, and cubic constants - require tuning. Although these approaches offer faster solutions compared to the channel-by-channel approach, they may not be rapid enough for online provisioning in dynamic planning scenarios.
All these proposed approaches involve analyzing an extensive search space \cite{Correia_PowerControl, Luo_22_optic_express,RanjbarECOC22022}. Furthermore, as demonstrated in \cite{Ives_JLT}, for C-band WDM systems, considering flat received power at the end of the span as an objective function results in higher link capacity compared to using flat GSNR as the objective function. 

Additionally, the authors in \cite{Lasagni_ECOC2021} and \cite{Souza_Genetic} have shown that minimum GSNR maximization or optimizing GSNR flatness as an objective function does not maximize the system capacity. 
Therefore, in the context of this paper, we have opted to focus on maintaining a uniform power level after each span while considering the system capacity maximization. To achieve this, in contrast to the (tilt, offset) approach, we introduce a hyper-accelerated power optimization (HPO) method, where we seek to determine the optimal flat launch/received power by solving the Raman coupled differential equations in forward/backward mode. The runtime for this optimization averages approximately 3 [sec] for span lengths ranging from 50 to 100 km.  
As we will demonstrate, embracing this approach not only results in the rapid resolution of power optimization, measured in a few seconds rather than several minutes \cite{Correia_PowerControl} or even hours \cite{Semrau_PowerOptimization}, but it also contributes to an improvement in the average GSNR per span, taking into consideration the overall span capacity. 
Moreover, it increases the utilization of higher modulation formats throughout the network. Additionally, in practical terms, the OSNR is a parameter that can be easily monitored, measured, and integrated into network management systems. 
Therefore, the uniformity of the OSNR, rather than the GSNR, becomes particularly significant, especially when different channels or lightpaths exhibit varying GSNR values due to the disparate routes they follow \cite{MultiBand_Amplifier_JLT}. 

Various approaches to static \cite{MehrabiJLT2021}, semi-static \cite{JanaJOCN022QoT}, and dynamic \cite{SamboJLT2020} RMBS(W)A have been explored. In static planning, the traffic matrix is known in advance, and service lifecycles are considered infinite. Semi-static planning involves unknown traffic matrices but assumes infinite service lifecycles, while dynamic planning deals with unpredictable traffic matrices and lifecycles. It is worth noting that semi-static network planning is the practical case of backbone networks \cite{LordJLT2015}. Therefore, it has been considered in this study. The routing in optical network planning typically relies on the k-shortest path first (K-SPF) algorithm. This involves selecting the shortest path (based on distance, number of optical hops, or multiplexing sections) with sufficient available spectrum resources as the primary choice. However, some studies have explored alternative criteria such as maximizing GSNR \cite{Correia_PowerControl, JanaJOCN022QoT} or minimizing the cost of the load balancing factor \cite{ArpanaeiJOCN2020QoT} when selecting paths. Additionally, distance-adaptive network planning \cite{MehrabiJLT2021, commletter_distanc, Re_Verification2022JLT} has been explored, with some studies considering worst-case channel scenarios \cite{SamboJLT2020, Correia_PowerControl} for modulation cardinality index selection, determining the line card interfaces' (LCIs) bit rate capability. In the context of this article, we assume to transmit signal with quadrature amplitude modulation (QAM) employing probabilistic shaping. While recent works have addressed both fixed-grid \cite{Correia_PowerControl} and flexi-grid \cite{MehrabiJLT2021} MB-EONs, the latter pose practical challenges due to ISRS effects. Despite efforts to apply flexi-grid EONs, developing a closed formula to estimate power profiles, even for LC-band scenarios, remains unfeasible \cite{zefreh2020realtime}. Some authors have focused on QoT-aware network planning \cite{MehrabiJLT2021,JanaJOCN022QoT} but have overlooked the amplified spontaneous emission (ASE)-shaped noise filler for idle channels \cite{InfineraCL}, resulting in unaddressed power profile changes and QoT degradation in established lightpaths.

Furthermore, optical layer grooming can be achieved through transponders or Flexponders. 
In transponder grooming \cite{Correia_PowerControl,JanaJOCN022QoT}, each client card interface is connected to an LCI, with IP flows groomed from the client card to the corresponding LCI. Modern modems operate based on the Flexponder concept, allowing low-bit-rate IP flows between a source-destination to be groomed at any LCI with sufficient capacity and the same source-destination. As the GSNR of each channel varies, the modulation cardinality of each LCI differs based on the frequency work point. Wavelength-dependent QoT considerations, as discussed in this paper, can reduce blocking probability and result in cost savings through optimized LCI usage. 
% \AntoComm{actually, modern transceiver can change the symbol rate as well}
Therefore, this manuscript considers real-world MB-EONs in which the bit rate variable LCIs are achieved by adjusting the modulation cardinality.
The main contributions of this manuscript can be summarized as follows:
1) a hyper-accelerated power optimization approach for uniform and pre-tilt launch power profiles, termed flat launched power (FLP) and flat received power (FRP), in MB-EONs.
2) Given the non-flat GSNR profile of optimum MB-EONs, the cross-layer design presents numerous complexities. Hence, we propose a wavelength-dependent QoT-aware modulation cardinality index selection and spectrum assignment based on a Flexponder.
3) We conduct exhaustive simulations across various span lengths and network configurations to demonstrate the efficacy of the proposed HPO algorithm. 4) Fourthly, we compare the performance of local optimization versus global optimization to assess the algorithm's effectiveness.
We provide an overview of the works published in the literature in Table \ref{tab_state_of_the_art}.

% \AntoComm{Do you really need to have SPM, XPM defined here?}  SPM: Self-Phase Modulation, XPM: Cross-Phase Modulation, FRP: Flat Received Power
\begin{table}[!t]
\centering
\caption{Summary of the Related Works.$\alpha(f)$ is the fiber loss coefficient. CFM: Closed-form model.}
\label{tab_state_of_the_art}
\begin{tabular}{|c|ccccccccc|}
\hline
\multirow{2}{*}{Reference} & \multicolumn{9}{c|}{QoT-Aware Cross-Layer Design of MB-EONs}                                   \\ \cline{2-10} 
                           & \multicolumn{1}{c|}{\rotatebox[origin=c]{90}{ISRS effect on NLI (SPM+XPM)}} & \multicolumn{1}{c|}{\rotatebox[origin=c]{90}{ISRS effect on ASE}} & \multicolumn{1}{l|}{\rotatebox[origin=c]{90}{ISRS effect on $\alpha(f)$}} & \multicolumn{1}{c|}{\rotatebox[origin=c]{90}{\textbf{Optimum FRP}}}  & \multicolumn{1}{c|}{\rotatebox[origin=c]{90}{ASE-shaped Noise Filler}} & \multicolumn{1}{c|}{\rotatebox[origin=c]{90}{\textbf{ML-based GN/EGN CFM\cite{RanjbarECOC22022,poggiolini2018generalized,zefreh2020realtime}}}}          & \multicolumn{1}{c|}{\rotatebox[origin=c]{90}{\textbf{Flexponder   Grooming}}} & \multicolumn{1}{c|}{\rotatebox[origin=c]{90}{LCS-band}}          & \rotatebox[origin=c]{90}{Wavelength-Dependent }           \\ \hline
\cite{SamboJLT2020}                      & \multicolumn{1}{c|}{\checkmark}                   & \multicolumn{1}{c|}{\checkmark}           & \multicolumn{1}{c|}{\checkmark}     & \multicolumn{1}{c|}{$\times$} & \multicolumn{1}{c|}{\checkmark}               & \multicolumn{1}{c|}{$\times$} & \multicolumn{1}{c|}{$\times$}          & \multicolumn{1}{c|}{\checkmark}   & $\times$ \\ \hline
\cite{MehrabiJLT2021}                    & \multicolumn{1}{c|}{\checkmark}                   & \multicolumn{1}{c|}{$\times$}         & \multicolumn{1}{c|}{$\times$}   & \multicolumn{1}{c|}{$\times$} & \multicolumn{1}{c|}{$\times$}             & \multicolumn{1}{c|}{$\times$} & \multicolumn{1}{c|}{$\times$}          & \multicolumn{1}{c|}{$\times$} & $\times$ \\ \hline
\cite{Correia_PowerControl}     & \multicolumn{1}{c|}{\checkmark}                   & \multicolumn{1}{c|}{\checkmark}           & \multicolumn{1}{c|}{\checkmark}     & \multicolumn{1}{c|}{$\times$}   & \multicolumn{1}{c|}{\checkmark}               & \multicolumn{1}{c|}{$\times$} & \multicolumn{1}{c|}{$\times$}          & \multicolumn{1}{c|}{\checkmark}   & $\times$ \\ \hline
\cite{JanaJOCN022QoT}                       & \multicolumn{1}{c|}{$\times$}                 & \multicolumn{1}{c|}{\checkmark}           & \multicolumn{1}{c|}{$\times$}   & \multicolumn{1}{c|}{$\times$} & \multicolumn{1}{c|}{$\times$}             & \multicolumn{1}{c|}{$\times$} & \multicolumn{1}{c|}{$\times$}          & \multicolumn{1}{c|}{$\times$} & \checkmark   \\ \hline
\cite{Re_Verification2022JLT}        & \multicolumn{1}{c|}{\checkmark}                   & \multicolumn{1}{c|}{$\times$}         & \multicolumn{1}{c|}{$\times$}   & \multicolumn{1}{c|}{$\times$} & \multicolumn{1}{c|}{$\times$}             & \multicolumn{1}{c|}{$\times$} & \multicolumn{1}{c|}{$\times$}          & \multicolumn{1}{c|}{$\times$} & $\times$ \\ \hline
\cite{ThreeStageSong2023}        & \multicolumn{1}{c|}{\checkmark}                   & \multicolumn{1}{c|}{\checkmark}           & \multicolumn{1}{c|}{$\times$}   & \multicolumn{1}{c|}{$\times$}   & \multicolumn{1}{c|}{$\times$}             & \multicolumn{1}{c|}{$\times$}   & \multicolumn{1}{c|}{$\times$}          & \multicolumn{1}{c|}{$\times$} & $\times$ \\ \hline
\cite{commletter_distanc}                     & \multicolumn{1}{c|}{\checkmark}                   & \multicolumn{1}{c|}{$\times$}         & \multicolumn{1}{c|}{$\times$}   & \multicolumn{1}{c|}{$\times$} & \multicolumn{1}{c|}{$\times$}             & \multicolumn{1}{c|}{$\times$} & \multicolumn{1}{c|}{$\times$}          & \multicolumn{1}{c|}{\checkmark}   & $\times$ \\ \hline
\textbf{This study}     & \multicolumn{1}{c|}{\checkmark}                   & \multicolumn{1}{c|}{\checkmark}           & \multicolumn{1}{c|}{\checkmark}     & \multicolumn{1}{c|}{\checkmark}   & \multicolumn{1}{c|}{\checkmark}               & \multicolumn{1}{c|}{\checkmark}   & \multicolumn{1}{c|}{\checkmark}            & \multicolumn{1}{c|}{\checkmark}   & \checkmark   \\ \hline
\end{tabular}
\end{table}

The rest of this paper is structured as follows: Section \ref{sec_II_QoT}
provides a summary and analysis of the QoT estimator and physical layer modeling. 
Section \ref{sec_III_nodeAndNetArchi} describes the node and network architecture compatible with the proposed cross-layer design methodology. 
Section \ref{sec_IV_HPO} introduces the power optimization algorithms proposed in this paper, including the FLP and FRP algorithms. 
Section \ref{sec_V_Sim} evaluates the performance of the proposed solution. 
Finally, Section \ref{sec_VI_conclusion} presents the conclusions of this paper.

\section{QoT Estimation Tool and Physical Layer Modeling}\label{sec_II_QoT}
% Before delving into a discussion of our algorithm for power optimization of MB-EONs, let us first begin with 
This section describes the physical layer modeling and QoT estimation, followed by an examination of our methodology. 
In this paper, we specifically address the GSNR when referring to the QoT, accounting for linear and non-linear noise components.  

\subsection{End-to-End QoT Estimator}\label{subsec:QoT}

\begin{figure*}[!t]
\centering
\includegraphics[width=1\linewidth]{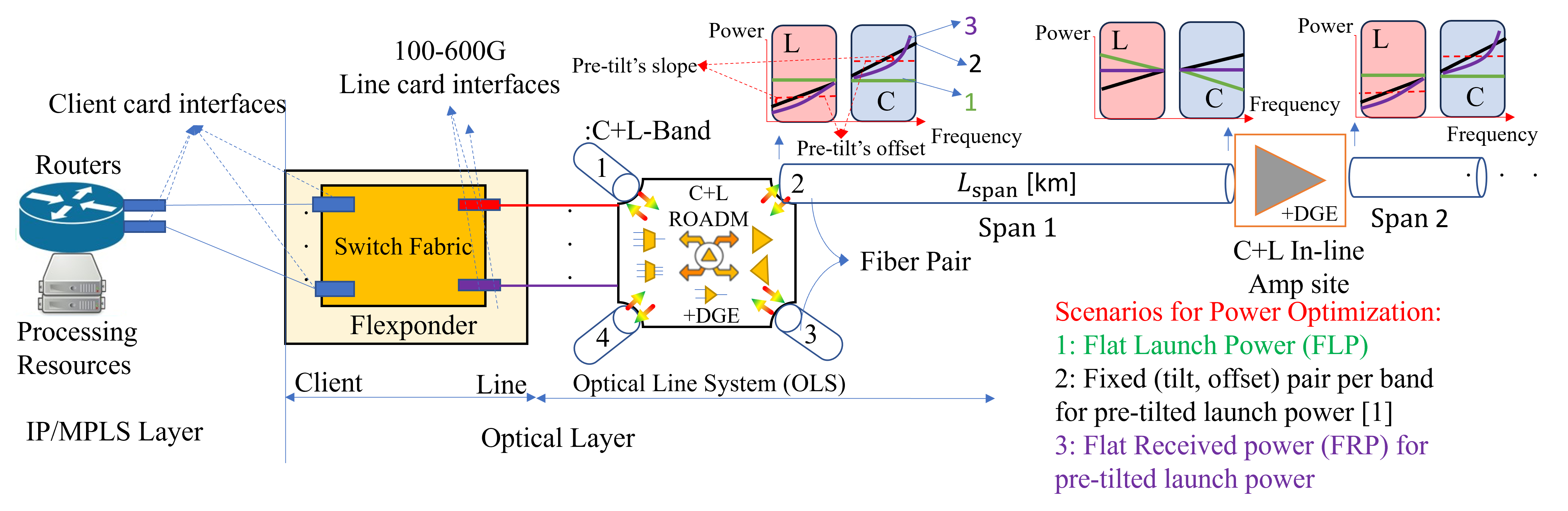}
\caption{A cross-layer architecture is proposed for MB-EONs, focusing on illustrating C+L-band systems for simplicity.}
\label{Fig:SystemModel_PowerOpt}
\end{figure*}

The Gaussian-noise (GN) model \cite{Poggiolini2014} is utilized to examine the effects of both linear factors, such as loss and chromatic dispersion, and non-linear interference (NLI) effects such as self-phase modulation (SPM), cross-phase modulation (XPM), multi-channel interference (MCI), and ISRS, on the optical signals' amplitude and/or phase modulation as they propagate through an optical fiber medium.
Since we have considered the standard single-mode fiber (SSMF) and we are not working in zero or very low fiber dispersion regimes, the MCI is negligible \cite{zefreh2020realtime}. Consequently, the computation of GSNR for a specific channel $i$ within span $s$ on link $l$ is achieved using  \eqref{eq:GSNR_Span} and \eqref{eq:P_start}:
% \begin{equation}\label{eq:GSNR_Span}
% 	GSNR^{l,s,i} \cong \frac{P^{l,s,i}_{\text{tx}}}{P^{l,s,i}_{\text{ASE}}+P^{l,s,i}_{\text{NLI}}}\text{,} 
% \end{equation}
\begin{equation}\label{eq:GSNR_Span}
	GSNR^{l,s,i} \cong \frac{P}{P^{l,s,i}_{\text{ASE}}+P^{l,s,i}_{\text{NLI}}}\text{,} 
\end{equation}

\begin{equation}\label{eq:P_start}
	P =\begin{cases}
	P^{l,s+1,i}{(z=0)}\,\,\,\,; & \text{ if } s<N_{\text{s}}^{l}\,\,, \\
 	P^{l+1,1,i}{(z=0)}\,\,\,\,; & \text{ if } s=N_{\text{s}}^{l}, l<N_{\text{L}}\,\,, \\
	P^{l,s,i}{(z=0)}\,\,\,\,;& \text{ if } s=N_{\text{s}}^{N_{\text{L}}},
	\end{cases}
\end{equation}
% \begin{equation}\label{eq:Operator_Star}
% 	P =\begin{cases}
% 	P^{l,s+1,i}_{\text{tx}}(z=0)\,\,\,\,; & \text{ if } s<N_s^l\,\, \text{and}\,\, \omega^{T^s,r',l'}=0, \\
% 	1\,\,\,\,;& \text{ if } \omega^{T^s,r,l}\,\text{or}\, \omega^{T^s,r',l'} =1 ,\\
% 	-1;& \text{ Otherwise }.
% 	\end{cases}
% \end{equation}
$N_\text{s}^{l}$ represents the number of spans in link $l$, while $N_\text{L}$ denotes the link number of the corresponding lightpath (LP). 
Utilizing the incoherent GN model for long enough spans, the value of the overall GSNR for a LP on channel $i$ can be obtained from \eqref{eq:GSNR_total}:
\begin{multline}\label{eq:GSNR_total}
	GSNR^{i}_{\text{LP}}|_{_{\text{dB}}} = \\10 \log_{10} \left[\left(\sum_{l=1}^{N_{\text{L}}}\sum_{s=1}^{N_{\text{s}}^l}\frac{1}{GSNR^{l,s,i}}+SNR_{\text{TRx}}^{-1}\right)^{-1}
 \right] \\-  SNR_{\text{Pen}_{\text{fil}}}|_{_{\text{dB}}} - SNR_{\text{Pen}_{\text{age}}}|_{_{\text{dB}}},
\end{multline}
where various parameters contribute, such as the power evolution profile (PEP) of each span ($P^{l,s,i}(z), 0\leq z \leq L_\text{s}^{l,s}$), the length of span $s$ in link $l$ ($L_\text{s}^{l,s}$), the noise power caused by the optical amplifier ($P^{l,s,i}_{\text{ASE}}= n_{\text{F}}hf^i(G^{l,s,i}-1)R^{i}_{\text{s}}$), and the noise power stemming from NLI ($P^{l,s,i}_{\text{NLI}}$), including SPM, XPM, and ISRS. Additionally, $SNR_{\text{TRx}}$, $SNR_{\text{Pen}_{\text{fil}}}$, and $SNR_{\text{Pen}_{\text{age}}}$ represent the transceiver SNR, SNR penalty due to wavelength selective switches (WSSs) filtering, and SNR margin due to aging, respectively \cite{Buglia2022, Sequeira2018, Pedro2022}.
Furthermore, the channel bandwidth ($B_{\text{ch}}^{i}$) and bit rate ($R_{\text{ch}}^{i,m}$) of each channel with modulation cardinality $m$ are calculated from $B_{\text{ch}}^{i} = \left \lceil \frac{R^{i}_{\text{s}}(1+\rho^{i})} {B_{\text{Base}}} \right \rceil B_{\text{Base}}$, and $R_{\text{ch}}^{i,m} = m  R^{i}_{\text{s}}(1+\rho^{i})(1-\theta)$, respectively. Here, parameters like the symbol rate of the channel ($R^{i}_{\text{s}}$), the roll-off factor ($\rho^{i}$), the forward error correction (FEC) overhead ($\theta$), and the bandwidth of a base frequency slot ($B_{\text{Base}}$) are involved.
% \begin{equation}\label{eq:BW} 
% B_{\text{ch}}^{l,i} = \lceil \frac{R^{l,i}_{\text{s}}(1+\rho^{l,i})} {B_{\text{Base}}} \rceil \times B_{\text{Base}}, 
% \end{equation}
% \begin{equation}\label{eq:R_Channel} 
% R_{\text{ch}}^{l,i,m} = m \times R^{l,i}_{\text{s}} \times (1+\rho^{l,i}) \times (1-\theta), 
% \end{equation}
% Indeed, $ P^{l,s,i}_{\text{tx}}(z=L)/P^{l,s,i}_{\text{tx}}(z=0)$ is the loss evolution profile along the fiber that shows the ISRS effect, i.e., depleting power from higher frequency channels to the lower frequency channels \cite{Semrau_NLI}
Moreover, the GSNR threshold for each modulation format level depends on the pre-FEC bit error rate (BER) and can be determined using (7) in \cite{6DMANJOCN2024}.
It is assumed that the boosters' gains at the add and pass-through directions are $G^{l,s,i} = 20$ dB, and the pre-amplifier can completely compensate for the fiber loss of the link and the QoT degradation caused by ISRS. Thus, for the pre-amplifiers and in-line amplifiers, the following expressions can be written: $G^{l,s,i} =  P/P^{l,s,i}(z=L_\text{s}^{l,s})$, where $P$ (see \eqref{eq:P_start}) and $ P^{l,s,i}(z=L_\text{s}^{l,s})$ are the powers of channel $i$ just after and before the corresponding amplifier, respectively. For instance, consider the scenario of flat launch power (FLP) in span 1, as illustrated in Fig. \ref{Fig:SystemModel_PowerOpt}. To create a practical model for the optical line system penalties, we applied a range of factors: randomly assigning connector losses between 0.2 to 0.5 dB, factoring in ROADMs' polarization dependent loss at 0.5 dB per node along the lightpath, and randomly accounting for splice losses in the range of 0.01 to 0.06 dB/km. The average length of the fusion splicing sections is considered 2 km \cite{Zhang_MBEON_800G}.
\subsection{Physical Layer Model}\label{subsec:QoT2}
Acquiring precise values for all parameters in GSNR calculation is often complex and sometimes unfeasible. Nevertheless, through the application of state-of-the-art telemetry and AI-based approaches \cite{Telemetri2021machine, Telemetri2021monitoring}, we can characterize physical layer parameters such as the noise figure of the amplifiers, ROADM's filtering penalty, and $SNR_\text{TRx}$. Assuming we possess acceptable knowledge of these physical layer parameters, the most challenging aspect of GSNR estimation for MB-EONs lies in estimating $P^{l,s,i}_{\text{NLI}}$ and $P^{l,s,i}_{\text{ASE}}$, with the ISRS effects playing a predominant role.
However, while the NLI models in the time-frequency domain offer the highest accuracy, they rely on solving complex integrals and are unsuitable for online or offline network planning tools \cite{Poggiolini2014}. The computational time needed by models like the split-step Fourier method, integral-based GN model, and enhanced GN model (EGN) is excessively high \cite{Poggiolini2017}. Furthermore, they are too complex to adequately account for the add/drop effect modeling in network-wide level studies. Consequently, over the past several years, several closed-form transmission models (CFMs) (e.g., \cite{Semrau2019}) and semi-CFMs (e.g., \cite{zefreh2020realtime,Buglia2023}) have been developed to estimate NLI. CFMs provide closed-form formulas for PEP and NLI but rely on specific assumptions for each model, e.g., triangular shape for Raman gain profile \cite{Semrau2019}. If the system model does not align with these assumptions, the model's accuracy diminishes. Semi-CFMs, on the other hand, calculate PEP and loss coefficients using fitting approaches, allowing for flexibility in ignoring certain assumptions. Additionally, the generalized GN model (GGN) is a well-known GN integral-based QoT estimator widely used in DWDM systems, offering acceptable accuracy but lacking modulation format correction terms crucial for MB-EONs \cite{GGNModel}.
Four fast CFMs have recently surfaced in the literature, as discussed in \cite{Semrau2019, Dimitris, Souza_JOCN_fastQoT, LasagniJLT2023}. The authors of \cite{Souza_JOCN_fastQoT} compared the models proposed in \cite{Semrau2019, Dimitris} with the GGN model \cite{GGNModel}, which is utilized in GNPy \cite{CurriGNPy}. They found that the model presented in \cite{Semrau2019} demonstrates the highest accuracy for LCS1-band scenarios, particularly when additional correction forms are incorporated. However, their investigation primarily focused on Gaussian-shaped signals. Conversely, the authors of \cite{LasagniJLT2023} introduced a CFM that considers Raman windowing sweeping across the frequency axis to enhance the accuracy of the model proposed in \cite{Semrau2019}, specifically for LC-band scenarios.
Two semi-CFMs introduced in \cite{zefreh2020realtime,Buglia2023} offer adequate accuracy for EONs beyond the C+L+S1-band. In this paper, we employ the model in \cite{zefreh2020realtime,RanjbarECOC22022} that uses the most advanced techniques, which is a machine learning (ML)-based GN/EGN model that has been validated through both the split-step Fourier method and experimental testing \cite{Yanchao_ECOC_2023,Yanchao_IPC_2023}. In addition, it incorporates essential features such as dispersion and modulation format correction terms, striking a fine balance between accuracy and speed. To estimate the $GSNR^{l,s,i}$ of each span, we follow the methodologies outlined in \cite{RanjbarECOC22022}, utilizing \eqref{eq:GSNR_Span}-\eqref{eq_gamma}. The process involves the following steps:
\begin{figure}[!t]
\centering
\includegraphics[width=1\linewidth]{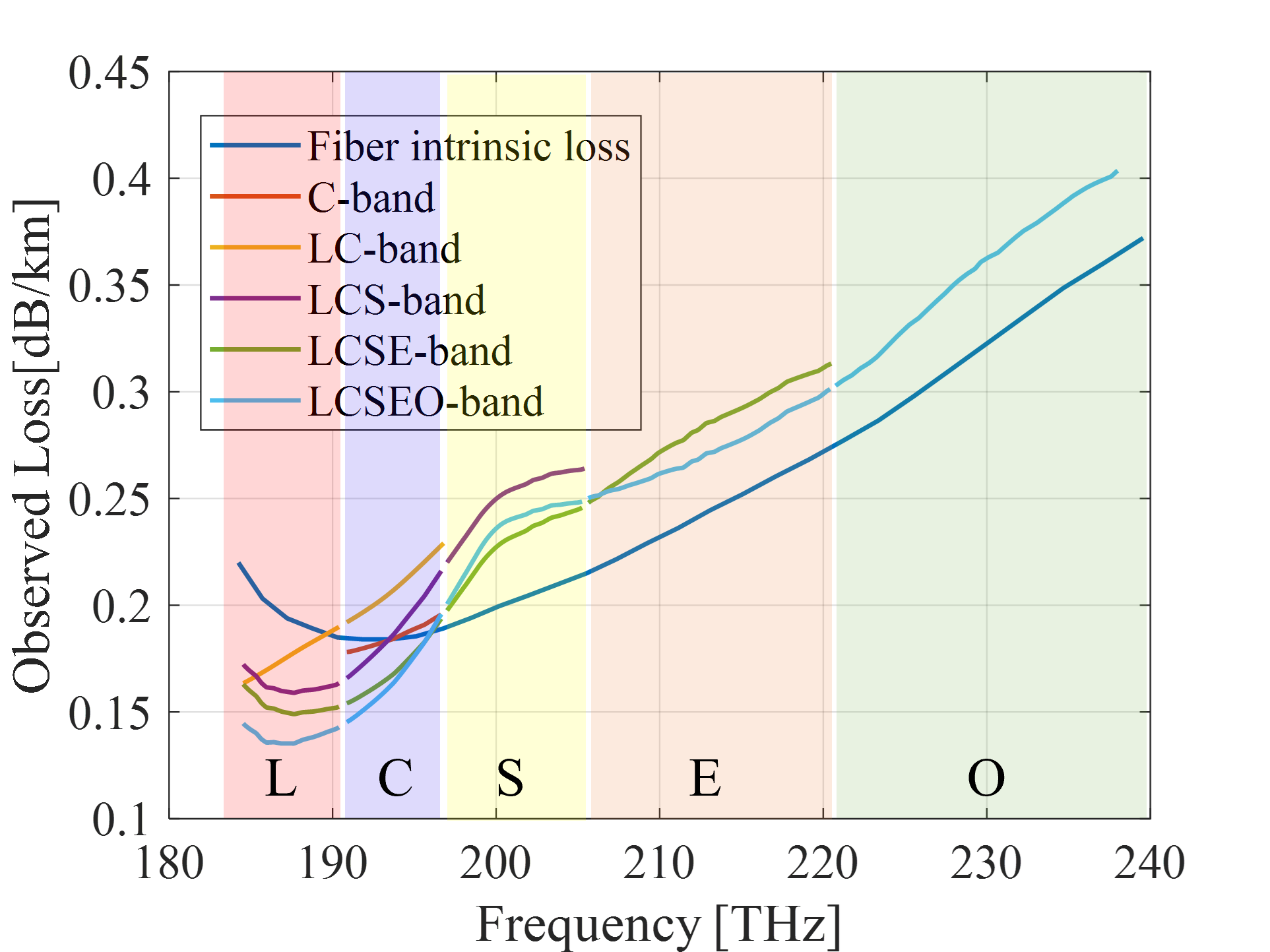}
\caption{ISRS effect on the fiber intrinsic loss for different multi-band systems. Span length = 70 km, launch power = 0 dBm, Fiber type: standard single mode fiber zero-peak water.}
\label{fig:loss_fitting}
\end{figure}
% This yields values for $\alpha^{l,s,i}_{0}$, $\alpha^{l,s,i}_{1}(f)$, and $\sigma(f)$.
% \begin{enumerate}

\noindent
\textbf{\textit{\underline{Step 1:}}} Numerically calculating the PEP by solving a system of coupled differential equations, i.e., \eqref{eq:ODE} detailed in Section \ref{sec_IV_HPO}. This necessitates the launch power profile, fiber loss coefficient profile, and the ISRS gain profile function. Notably, the ISRS gain is contingent upon factors such as the pump channel frequency, the discrepancy between pump and signal channels, and fiber physical parameters like effective area, diffraction coefficient, and numerical aperture value.
\textbf{\textit{\underline{Step 2:}}} Estimating auxiliary loss coefficients profiles, i.e., $\alpha_0(f),\, \alpha_1(f), \, \text{and} \, \sigma(f)$ by fitting the power evolution profile obtained from \textbf{step 1} and approximately closed-form formula, (13) in \cite{zefreh2020realtime}. The inspired actual frequency-depended fiber loss can be loosely modeled based on \eqref{eq_alpha}.

\begin{equation}\label{eq_alpha}
   \alpha(z,f_i) = \alpha_0(f_i)+\alpha_1(f_i)\exp\{-\sigma(f_i)z\},
\end{equation}   
where $z$ is the signal propagation distance, and the index $i$ to denote the channel's frequency \(f_i\). Indeed, the interpretation of \eqref{eq_alpha} suggests that the observed loss coefficient in MB systems differs from the loss coefficient utilized in C-band systems. In this context, \(\alpha_0(f_i)\) represents the fiber loss in the absence of ISRS, while \(\alpha_1(f_i)\) quantifies the loss alteration attributable to ISRS at the onset of the span. Additionally, \(\sigma(f_i)\) characterizes the rate at which ISRS diminishes along the span with the decreasing optical power. Once these parameters are assigned, the NLI calculation becomes closed-form. However, assigning auxiliary loss coefficient profiles is not closed-form, as doing it in full closed-form would introduce excessive errors. This step is the only non-closed-form, being a semi-CFM QoT estimation approach. Equations (30.1) and (30.2) in \cite{zefreh2020realtime} offer a closed-form best-fit for \(\alpha_1(f_i)\) and \(\alpha_0(f_i)\) given \(\sigma(f_i)\). Finally, optimizing numerically over \(\sigma(f_i)\) gives the overall best-fit for \(\alpha_1(f_i)\), \(\alpha_0(f_i)\), and \(\sigma(f_i)\). For illustration, Fig. \ref{fig:loss_fitting} presents the observed loss for a span length of 70 km in a fully loaded link with a flat launch power of 0 dBm. The results are obtained using the values outlined in Section \ref{sec_V_Sim} for all the necessary parameters. As shown, the observed loss curve differs from the expected loss curves due to ISRS effects across various scenarios of the MB-EONs. Additionally, the results indicate that these differences become more pronounced with each additional band.
\noindent
\textbf{\textit{\underline{Step 3:}}} Calculating $G^{l,s,i} =  P/P^{l,s,i}(z=L_\text{s}^{l,s})$ to derive $P^{l,s,i}_{\text{ASE}}$.
\noindent
\textbf{\textit{\underline{Step 4:}}} Computing the parameter $M$ from \eqref{eq_M}.
        \begin{equation}\label{eq_M}
   M = \mathrm{MAX}{\lfloor 10\times | 2\alpha_1(f_i)/\sigma(f_i)| \rfloor +1}.
    \end{equation} 
    \textbf{\textit{\underline{Step 5:}}} Determining the effective dispersion profile using \eqref{eq_beta}.
    \begin{multline}\label{eq_beta}
    \bar{\beta}_{2}^j = \beta_2 + \pi \beta_3(f_i+f_j-2f_0)+\frac{2\pi^2}{3}\times\\
 \beta_4[(f_i-f_0)^2+(f_i-f_0)(f_j-f_0)+(f_j-f_0)^2].
\end{multline}
Here, \(f_0\) denotes the frequency reference, which is associated with the wavelength 1550 nm, where \(\beta_2\), \(\beta_3\), and \(\beta_4\) are measured. 
\noindent
\textbf{\textit{\underline{Step 6:}}} Establishing the frequency-dependent non-linearity coefficient using \eqref{eq_gamma}.
\begin{equation}\label{eq_gamma}
 \gamma^{i,j} = \frac{2 \pi f_i}{c}\frac{2n_2}{A_{\text{eff}}(f_i)+A_{\text{eff}}(f_j)},
\end{equation}  
where $n_2$ is the nonlinear (Kerr) refractive index and $A_{\text{eff}}$ is the effective area. 
\noindent
\textbf{\textit{\underline{Step 7:}}} Computing $P^{l,s,i}_{\text{NLI}}$ based on (2) in \cite{RanjbarECOC22022}.
% \end{enumerate}
Therefore, by substituting the values of $P^{l,s,i}_{\text{NLI}}$ and $P^{l,s,i}_{\text{ASE}}$ for each span in a LP, the total end-to-end GSNR can be calculated using \eqref{eq:GSNR_total}.
\section{Node and Network Architectures, and Traffic Engineering}\label{sec_III_nodeAndNetArchi}
\begin{figure*}[!t]
\centering
\includegraphics[width=1\linewidth]{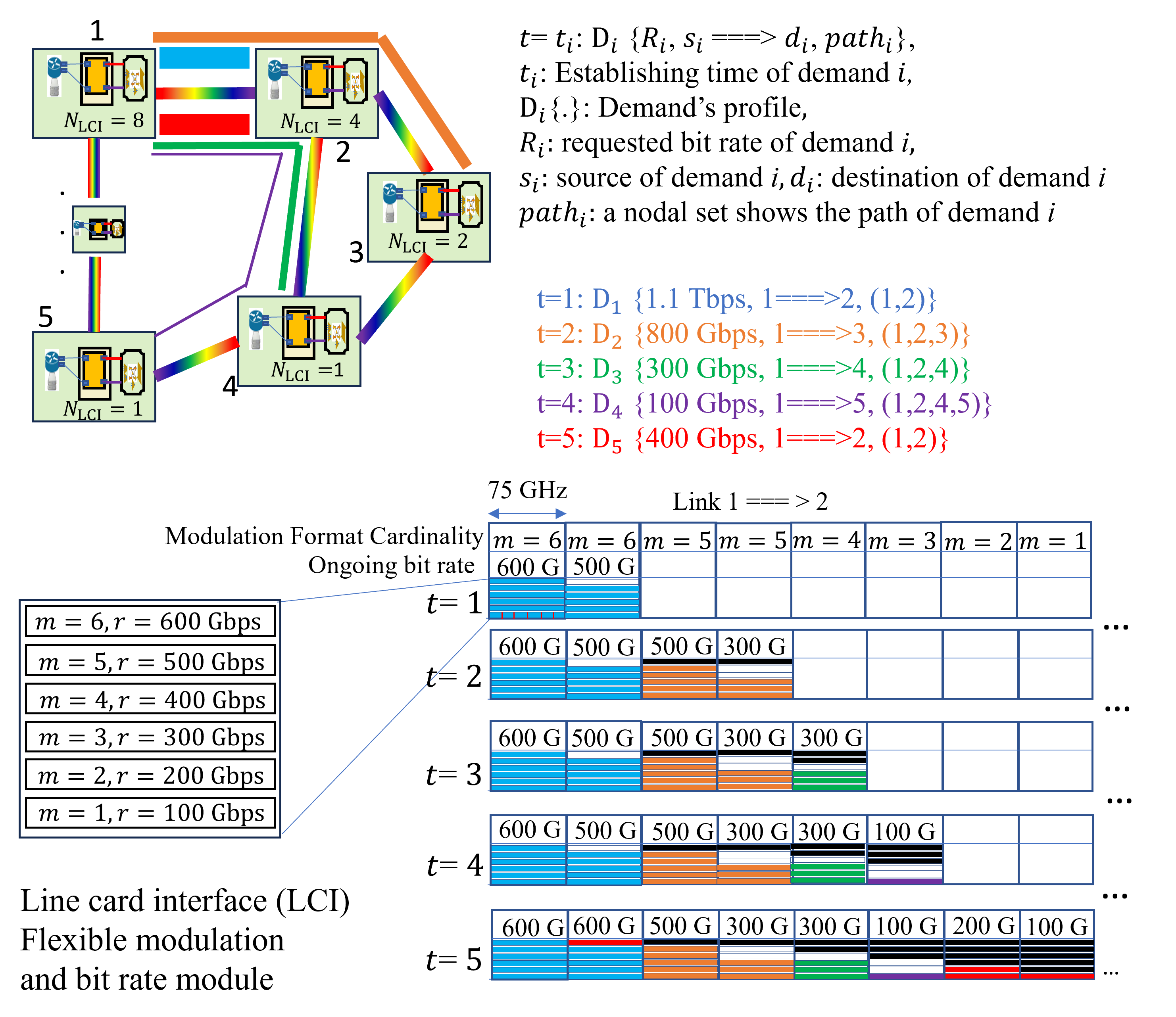}
\caption{Illustrative example for Flexponder grooming based on the wavelength-dependent QoT-aware cross-layer design. }
\label{fig:Grooming_Multiband}
\end{figure*}
In this section, we outline the nodal and network architecture for the next-generation MB-EONs, where the optimal power of each span can be swiftly adjusted based on physical layer variations throughout the network's lifecycle.
\subsection{Assumptions for Node Architecture}\label{subsec:NodeArc}
Each node is comprised of two layers as illustrated in Fig. \ref{Fig:SystemModel_PowerOpt}: the IP/MPLS layer, housing routers, switches, and processing resources, and the optical layer, consisting of electrical-optical equipment such as Flexponders and ROADMs. Flexponders include client grey card interfaces, switch fabrics (optical transport network, i.e., ITU-T G.709 or a Muxponder), and colored line card interfaces (LCIs). Fig. \ref{fig:Grooming_Multiband} depicts LCIs with flexible modulation formats and bit rates. Leveraging state-of-the-art technologies like probabilistic constellation shaping, super-fast digital signal processing ASIC, and soft decision (SD)-FEC, LCIs can operate within a range of 100 Gbps to 1.6 Tbps over varying distances, with modulation format and FEC adjustments. For the sake of simplicity, we have considered LCIs ranging from 100 Gbps to 600 Gbps. Consequently, the GSNR gauge profile can reflect real-time QoT, and the bit rate can adapt according to the modulation format of the LCIs. Hence, the software-defined networking (SDN) controller can dynamically adjust traffic grooming to allocate traffic to feasible LCIs. Client-side traffic from the IP/MPLS layer can be groomed electronically within the Flexponders and then forwarded to the LCIs (see Fig. \ref{Fig:SystemModel_PowerOpt}). 
The architecture features colorless-directionless, and contentionless ROADMs-on-blade \cite{6DMANJOCN2024}, which include boosters, pre-amplifiers, WSSs, and digital gain equalizers (DGEs) connected to fiber pairs. Each link (connecting two nodes) is divided into several spans (connecting cascaded in-line amplifier (ILA) sites), with the ILA site also incorporating DGEs to adjust launch power. It is worth noting that we assume the ILA site is equipped with gain and power controllers. Additionally, optical amplifiers within the ROADMs and ILAs for each band vary, requiring a multiplexer/demultiplexer before and after them. 
\subsection{Assumptions for Network Architecture and Traffic Engineering}\label{subsec:NetArc}
The network is assumed to be managed by a multi-layer SDN orchestrator capable of controlling both the optical and IP/MPLS layers. Consequently, when new demands arise, or the QoT of established LPs deteriorates due to hardware or software failures, a QoT-aware service provisioning procedure is initiated by the orchestrator and the relevant SDN agent controller in each layer. As depicted on the top of Fig. \ref{fig:Grooming_Multiband}, we consider a network scenario highlighting five nodes and demands. Let us imagine that at $t=1$, the links between nodes 1 and 2 ($1\Longleftrightarrow2$), ($2\Longleftrightarrow3$ and 4), and ($4\Longleftrightarrow5$) are idle. Each demand is characterized by a tuple $D_{i} = \{R_i, s_i\Longleftrightarrow d_i, path_{i}\}$ as shown on the top of Fig. \ref{fig:Grooming_Multiband}. As mentioned in the previous section, LCIs can adaptively operate within a flexible range of modulation formats and bit rates, such as 100-600 Gbps corresponding to modulation cardinality $m=\{1, 2, 3, 4, 5, 6\}$. Moreover, based on \cite{Curri_ASE_21}, we assume that the idle channels are occupied by ASE-shaped noise to ensure network consistency. Therefore, whenever malfunctions leading to QoT degradation of LPs occur in the passive or active equipment of the optical layer, the proposed QoT-aware service provisioning procedure can be executed to optimize the network performance. As illustrated at the bottom of Fig. \ref{fig:Grooming_Multiband}, the first two channels of link $1\Longleftrightarrow2$ are occupied by two LCIs belonging to an LP with a capacity of 1.1 Tbps and $path_i=1\Longleftrightarrow2$. For $D_i$, based on the QoT estimator tool described in Section. \ref{sec_II_QoT}, both allocated LCIs operate with $m=$6 but at different bit rates, i.e., 600 Gbps and 500 Gbps. This implies that the second LCI has 100 Gbps spare capacity for grooming upcoming demands with the same source and destination. Similar procedures are followed for the third and fourth demands established at $t$= 3 and 4. At $t$= 5, the traffic grooming of $D_5$ and $D_1$ occurs, and instead of using an additional LCI, we establish it only with two additional LCIs operating in channels 7 and 8. Therefore, we require 8, 4, 2, 1, and 1 LCIs in nodes 1, 2, 3, 4, and 5, respectively.          

\section{Proposed Hyper-accelerated Power Optimization Method}\label{sec_IV_HPO}

According to \cite{Poggiolini2014}, a sub-optimal power configuration for a Nyquist-WDM optical network can be achieved by determining the optimal power allocation for each span, a process referred to as local optimization-global optimization (LOGON), where "N" represents Nyquist. Specifically, in an optical network utilizing Nyquist WDM channels with homogeneous spans—having identical physical layer characteristics such as length, noise figure, non-linearity, loss, dispersion coefficients, and fully loaded link state—the optimal flat launch power for each span can be computed based on (83) in \cite{Poggiolini2014}. It is important to note that these parameters are assumed to be frequency-independent.

However, if these assumptions are violated, such as in multi-band systems where ISRS effects are considered and physical layer parameters become frequency-dependent, (82) in \cite{Poggiolini2014} cannot be directly utilized. Consequently, a closed-form solution for this optimization problem is not available. Hence, we propose a heuristic algorithm to determine the optimal flat launch/receive power per span. Although this approach may not guarantee optimality, it leverages the principles of LOGON to obtain a sub-optimal power configuration for the network. To do so, we try to maximize the total capacity (TC) \cite{Souza_cost24} of each span while the maximum launch power can be $P_{\text{max}}$. For the sake of simplicity, we assumed that the
input constellation has a Gaussian distribution in the calculation of TC \cite{Buglia2022,arpanaei2023launch}. 

\begin{equation}\label{eq_opt}
\begin{aligned}
\textsc{MAX} \quad & TC^{l,s} = 2\sum_{i=1}^{N_{\text{ch}}} R^{l,s,i}_{\text{s}}\log_2(1+GSNR^{l,s,i})\\
\textrm{s.t.} \quad &\textbf{ C1}: P^{l,s,i}\leq P_{\text{max}}, \forall i \in \mathcal{C}, s \in \mathcal{S}, l \in \mathcal{L}. \\
\end{aligned}
\end{equation}

The parameter $N_{\text{ch}}$ denotes the number of channels in each link. $\mathcal{C}$, $\mathcal{S}$, and $\mathcal{L}$ represent the sets of channels, spans, and links, respectively. As discussed in Section \ref{sec_I_intro}, we assess two HPO algorithms: flat launch power (FLP) and flat receive power (FRP). To accomplish this, we must calculate the PEP. In an MB-EON with $N_{\text{ch}}$ channels having center frequencies $f_1 < f_2 < \dots < f_{N_{\text{ch}}}$, the evolution of power over distance for each channel (or PEP) is governed by a system of coupled differential equations \cite{zefreh2020realtime}:
% \begin{equation}
%     \frac{\partial P(f_j,z)}{\partial z} = \kappa P(f_j,z)\left\{\sum_{i=1}^{N}\zeta\left(\frac{f_j}{f_i}\right)\\C_r(f_i-f_j)P(f_i,z)-\alpha(f_j)\right\}, j \in [1, N],
%     \label{eq:eq_1}
% \end{equation}
\begin{multline}\label{eq:ODE}
	\frac{\partial  P^{l,s,i}_{\text{tx}}(z)}{\partial z} = \kappa P^{l,s,i}_{\text{tx}}(z)\Big[\sum_{j=1}^{N_{\text{ch}}}\zeta\left(\frac{f_i}{f_j}\right)\\C_r(f_j,f_j-f_i)P(f_j,z)-\alpha(f_i)\Big], i \in [1, N_{\text{ch}}],
\end{multline}
where $z$ is the signal propagation distance, $\alpha(f_i)$ is the fiber attenuation at frequency $f_i$, $P^{l,s,i}_{\text{tx}}(z)$ is the power of the $i^{th}$ channel at distance $z$, and $\kappa$ is set to $+1$ for a signal propagating along the $+z$ direction (forward propagating), while $\kappa{=}-1$ for signals propagating in the $-z$ direction (backward propagating). $\zeta(x)$ returns $x$ for $x > 1$, 0 for $x = 0$, and 1 for $x < 1$. $C_r$ exhibits odd symmetry with respect to the frequency difference $\Delta f = f_j - f_i$. This variable characterizes the gain profile of the Raman effect within the fiber. It is contingent on the fiber's physical attributes, such as the Raman gain coefficient and the effective area of the fiber. In FLP mode, we set $\kappa=1$ and restrict the range of acceptable launch power values to ensure a practical amplifier gain and total output power while managing nonlinear Kerr effects like SPM and XPM. A viable approach to counteract the ISRS effect is to incorporate a tilt in the output spectrum of the optical amplifiers in use, i.e., FRP mode. This tilt ensures a nearly uniform input spectrum into the subsequent optical amplifier, thus maintaining a consistent  OSNR after each span and at the receivers. To achieve this, we set $\kappa=-1$ and again restrict the range of acceptable launch power values to ensure a practical amplifier gain and total output power while managing nonlinear Kerr effects. Our objective is to determine the highest possible span capacity under these conditions. Algorithm \ref{alg_HPO} outlines the proposed HPO methods, FLP and FRP. In FLP mode, we set $P_{\text{start}} = P_{\text{LOGO}}$ and $P_{\text{end}}= P_{\text{LOGO}}+3$, where $P_{\text{LOGO}}$ can be determined using (83) in \cite{Poggiolini2014}. In FRP mode, we initiate our search with a starting channel power after propagation over a span of length $L_{\text{span}}$, calculated as $P_{\text{start}} = P_{\text{FLP,Opt}} - \alpha_{\text{max}} L_{\text{span}}$, where $P_{\text{FLP,Opt}}$ represents the optimal FLP and $\alpha_{\text{max}}$ is the maximum attenuation across all channel frequencies. Additionally, $P_{\text{end}}= 0$.

\begin{algorithm}[!t]
\caption{: HPO}\label{alg_HPO}
\begin{algorithmic}[1]
\renewcommand{\algorithmicensure}{\textbf{Output:}}
\renewcommand{\algorithmicrequire}{\textbf{Input:}}
\REQUIRE Spans' physical layer parameters, e.g., length, loss, dispersion, non-linearity coefficients, etc., and initial values such as $P_\text{start},\,P_\text{end},P_\text{max} $
\ENSURE Optimum TC, Launched and Received PEPs ($PEP_{\text{L/R}}$) and the Corresponding GSNR Profiles 
\STATE \textit{Initialization}:
    \STATE  $Max\_TC \gets$ 0.
    \FOR{$P^i \in (P_\text{start}:0.1:P_\text{end})$}
            \STATE Calculate $TC(P^i)$ from \eqref{eq_opt}.
            \IF{MAX \{$PEP$\} $> P_{\text{max}}$}
                \STATE  Break the loop
            \ENDIF
            \IF{$TC(P^i) > Max\_TC$}
            \STATE $Max\_TC \gets TC(P^i)$, $PEP_{\text{L/R}}^* \gets PEP_{\text{L/R}}$,\\ $GSNR^*\gets GSNR$,   \COMMENT{L:Launch, R: Receive}
            \ELSE
            \STATE  Break the loops
            \ENDIF
    \ENDFOR
\RETURN $Max\_TC,\ PEP_{\text{L/R}}^*,\, GSNR^* $.
\end{algorithmic}
\end{algorithm}
\section{Simulation Results and Discussion}\label{sec_V_Sim}
In this section, we compare FLP HPO against FRP HPO at both the span level and network-wide level. In the span level analysis, various multi-band scenarios, including C-, CL-, and LCS-band, are considered for spans ranging from 50 km to 100 km. The statistical values of the GSNR and OSNR profiles are reported to illustrate the performance of both FLP and FRP in the span-level study. Next, we move on to the network-wide analysis, where we evaluate the performance across three different-sized networks: small-scale (Spain), medium-scale (Japan), and large-scale (United States of America) backbone networks. The parameters assessed include bandwidth blocking probability, GSNR, number of LCIs, and modulation cardinality. Finally, we compare the performance of LOGON versus GON strategies at the network-wide level.
\subsection{Physical Layer Parameters}\label{subsec:Physical_Layer_Parameters}
\begin{figure}[!t]
\centering
\includegraphics[width=1\linewidth]{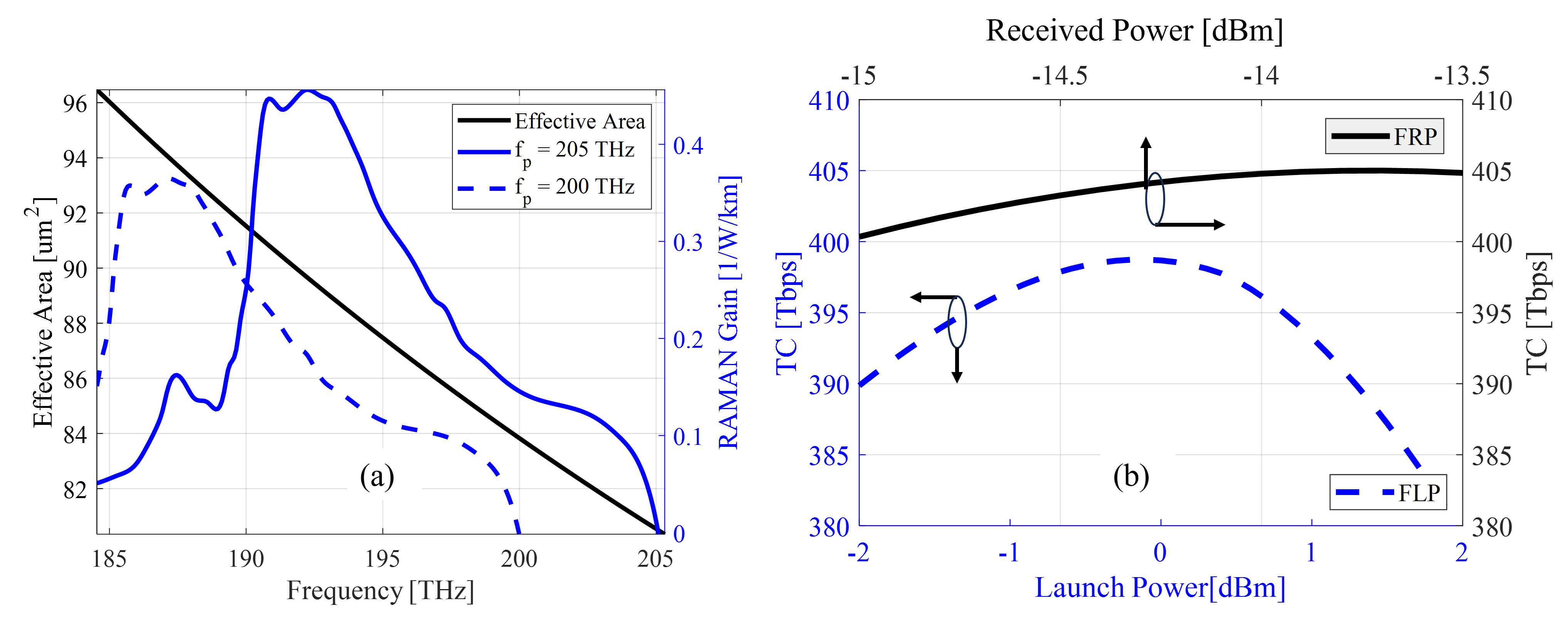}
\caption{(a) Effective area and Raman gain profile for two channels at 205 THz and 200 THz, (b) The hyper-accelerated power optimization algorithms, comparing flat launch power (FLP) with flat receive power (FRP), are evaluated for LCS-band scenario. The span length is set to $L_{\text{s}} = 70$ km.}
\label{fig:RamanGain_TCLCS70km}
\end{figure}

This paper considers an SSMF with zero water peak. Fig. \ref{fig:loss_fitting} depicts the fiber loss profile. The dispersion  is calculated based on \eqref{eq_beta} with $\beta_2 = -21.86 \times 10^{-27}$ [s$^2\cdot$ m$^{-1}$], $\beta_3 = 0.1331 \times 10^{-39}$ [s$^3\cdot$m$^{-1}$], and $\beta_4 = -2.7 \times 10^{-55}$ [s$^4\cdot$m$^{-1}$]. The measured effective area is illustrated in Fig. \ref{fig:RamanGain_TCLCS70km} , and $n_2 = 2.6 \times 10^{-20} $. The characterization of Raman gain profile $C_r(f_p,\Delta f)$ was initially conducted on a SSMF and subsequently adjusted in frequency \cite{RanjbarECOC22022}. Fig. \ref{fig:RamanGain_TCLCS70km} illustrates the plot of two instances of $C_r(f_p,\Delta f)$ for 200 THz and 205 THz pump frequencies. The assumption is made that the utilized link incorporates lumped amplifiers, including the following DFAs with their respective noise figures: Erbium DFA (EDFA) at 4.5 dB (C-band), EDFA at 5 dB (L-band), and Thulium DFA at 6 dB (S-band). Additionally, a total of 20 THz (6 + 6 + 8) is allocated for the LCS-band scenario \cite{cienawhitepaper}, with a 400 GHz gap between the bands. 
Moving forward, we assume that the fixed-grid and flexible bit-rate LCIs operate at 64 Gbaud, with a roll-off factor of 0.05, pre-FEC BER of $1.5 \times 10^{-2}$, and bit rates ranging from 100G to 600G. Therefore, according to \cite{6DMANJOCN2024}, the required GSNR, for $m$ = 1--6, is 3.45, 6.5, 8.4, 12.4, 16.5, and 19.3 dB, respectively. It is important to emphasize that these assumptions align with state-of-the-art technologies, such as SD-FEC  \cite{Alvarado2016}. Consequently, by allocating a 25\% overhead for SD-FEC, we can implement fixed-grid and flexible bit-rate LCIs ranging from 100 Gbps to 600 Gbps, whose modulation cardinality can be adjusted and controlled by an SDN controller. In this setup, the channel spacing is set at 75 GHz (6 $\times$ 12.5 GHz). 

\subsection{Span-level Study: FLP vs. FRP HPO}\label{subsec:FLPvsFRP}

\begin{figure}[!t]
\centering
\includegraphics[width=1\linewidth]{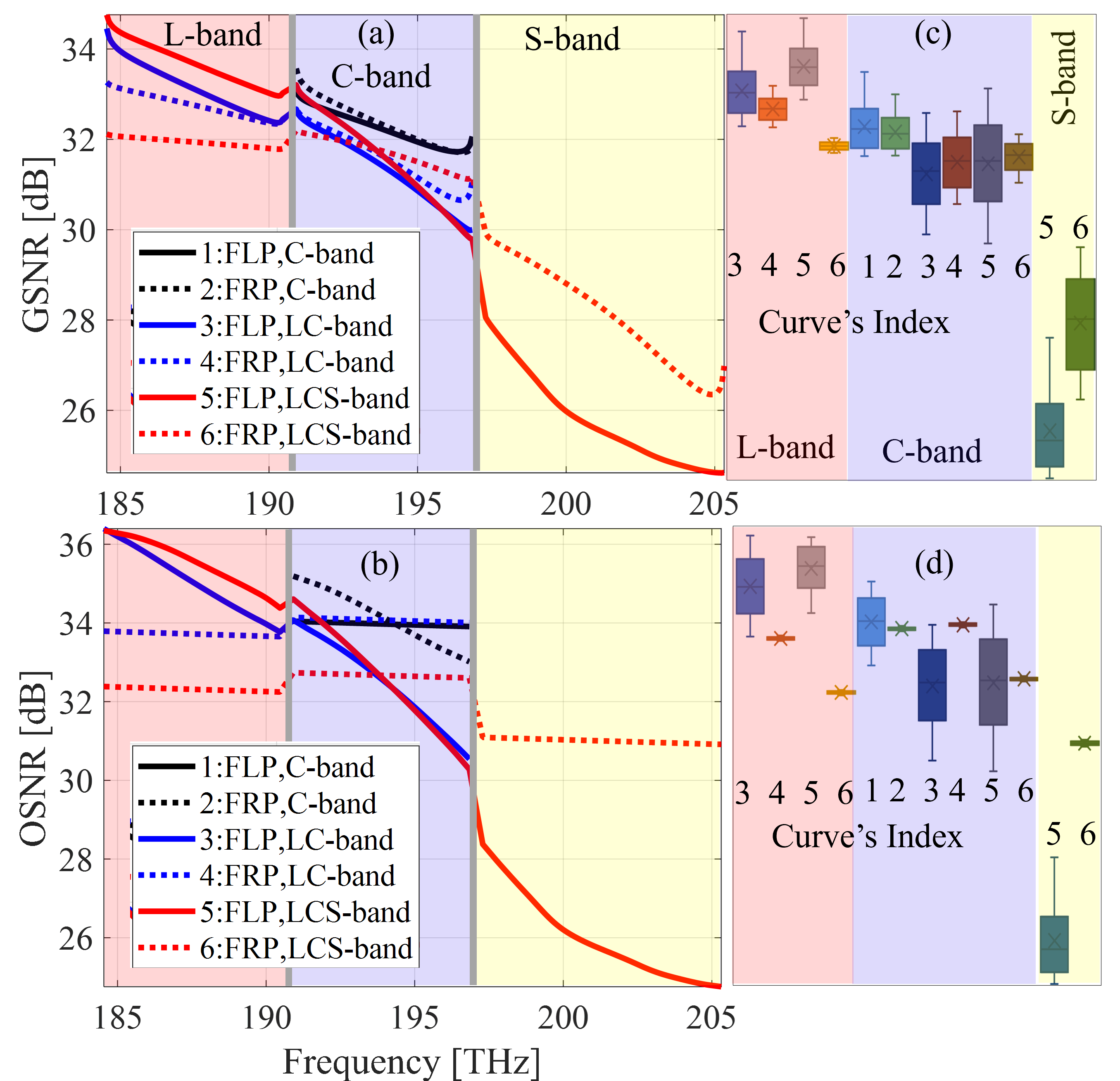}
\caption{The single span (a) GSNR profiles and (b) OSNR profiles for FLP and FRP C-band, LC-band, and LCS-band scenarios are depicted. The corresponding box plots are illustrated in (c) and (d) for L-, C-, and S-band. FLP: flat launch power FLP, FRP: flat receive power. The span length is set to $L_{\text{s}} = 70$ km.}
\label{fig:GSNR_OSNR_Freq_ALL_70km}
\end{figure}

\begin{figure}[!t]
\centering
\includegraphics[width=1\linewidth]{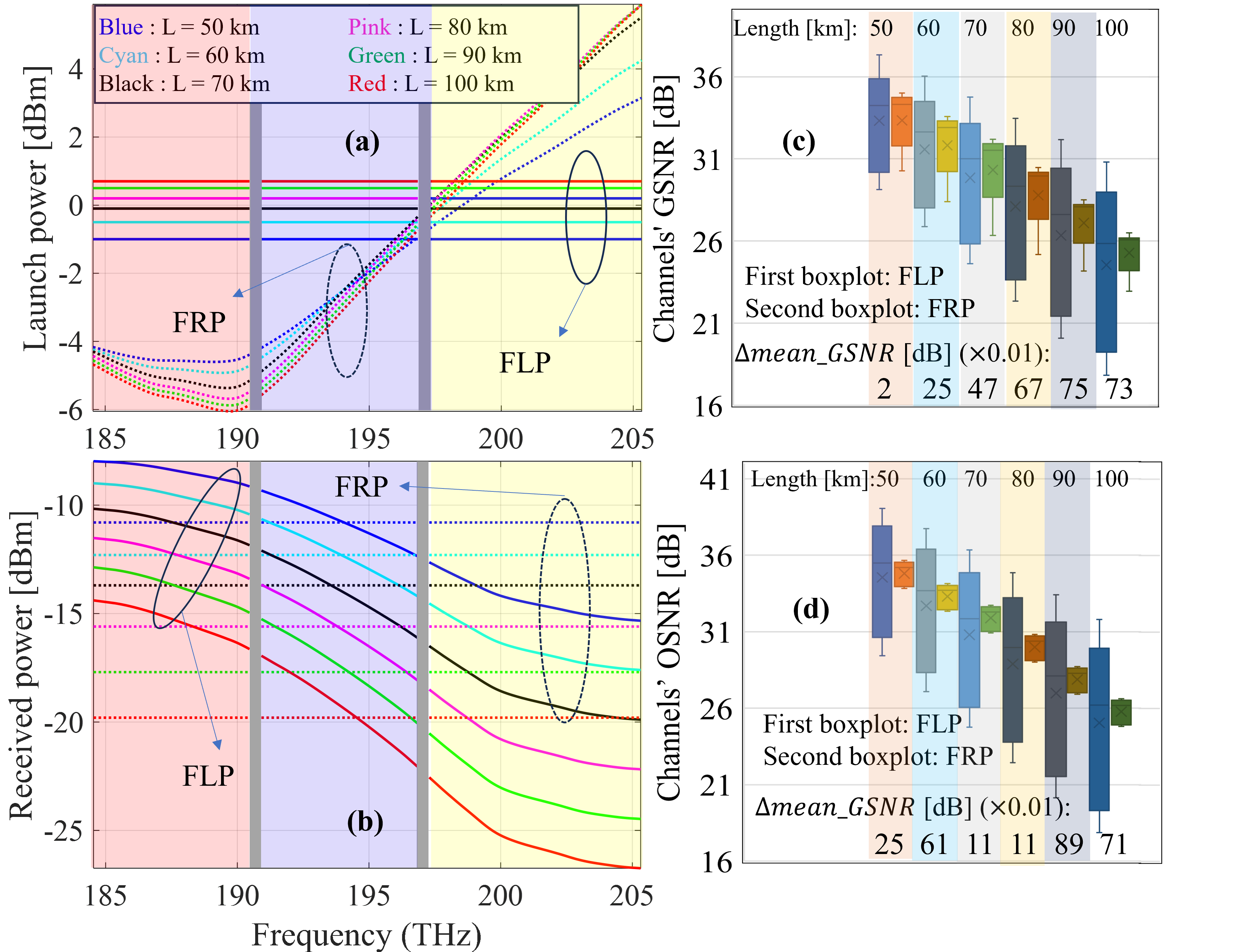}
\caption{Power, GSNR, and OSNR profiles comparing Flat Launch Power (FLP) and Flat Receive Power (FRP) for span lengths of 50-100 km. (a) Launch power profile, (b) Received power profile, (c) Box plots of GSNR profiles, and (d) Box plots of OSNR profiles for FLP and FRP.}
\label{fig:POWER_FREQ_50100}
\end{figure}

\begin{figure}[!t]
\centering
\includegraphics[width=0.7\linewidth]{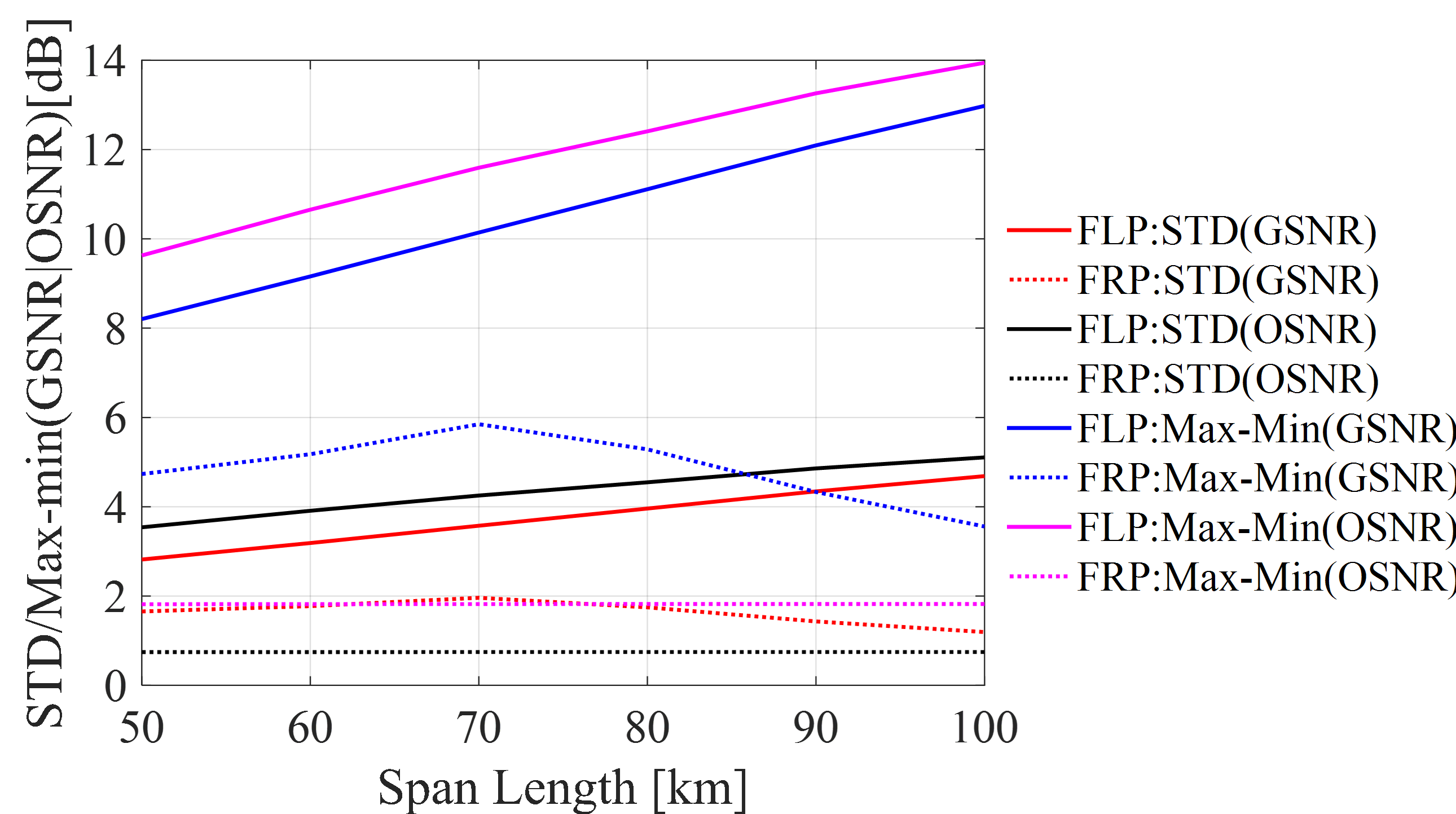}
\caption{Simulation results for the network.}
\label{fig:GSNR_OSNR_STD_MAXMIN_50100}
\end{figure}

The span-level analysis forms the foundation of this research. As the span-by-span power optimization is devised according to the LOGON strategy of the network, our initial focus is on demonstrating the performance of FRP and FLP at the span level. Given the constraints of the terrestrial backbone mesh network, where specifying long-haul paths with varying reach distances and considering DGE at specific ILAs is impractical, we assume that all ILAs are equipped with DGE.
According to Algorithm \ref{alg_HPO} outlined in Section \ref{sec_IV_HPO}, the FLP mode aims to find a flat launch power that maximizes the capacity of each span. Conversely, in the FRP mode, the objective is to find a flat receive power to maximize the capacity of the span. In the former approach, the PEP is computed by solving \eqref{eq:ODE} in the forward mode, while in the latter approach, it is derived by solving \eqref{eq:ODE} in the backward mode. As illustrated in Fig. \ref{fig:RamanGain_TCLCS70km}(b), the findings indicate an improvement of about 6.3 Tbps in TC over a 70 km span. In this scenario, we consider a flat received power across all bands. However, an alternative scenario could entail assigning a flat power per band based on the noise figure modification factor.
% Although we have opted not to present this result to conserve space, the findings indicate that maintaining a flat received power across all bands outperforms the scenario with flat received power per band.
% \begin{figure*}[!t]
% \centering
% \includegraphics[width=1\linewidth]{Figures/OCTvsPower_70km.png}
% \caption{The hyper-accelerated power optimization algorithms, comparing flat launch power (FLP) with flat receive power (FRP), are evaluated for (a) C-band, (b) LC-band, and (c) LCS-band scenarios. The span length is set to $L_{\text{s}} = 70$ km.}
% \label{fig:OCTvsPower_70km}
% \end{figure*}
To gain deeper insights into how FRP can enhance the TC of a span, we have assessed the GSNR and OSNR profiles along with their corresponding statistics, as depicted by the boxplots in Fig. \ref{fig:GSNR_OSNR_Freq_ALL_70km}. Once again, let us consider a span length of 70 km. As shown in Fig. \ref{fig:GSNR_OSNR_Freq_ALL_70km}(a), the GSNR improvement of channels in the C- and LC-band scenarios appears negligible. Particularly, the FRP introduces a tilt in the launch power, resulting in a seesaw effect. This means that to maintain a flat receive power, the launch power of higher-frequency channels must be higher than that of lower-frequency channels. Consequently, as demonstrated in \cite{arpanaei2023launch}, the GSNR of lower frequency channels (L-band) decreases, as indicated by the dashed blue curve in the L-band.
Conversely, the GSNR of C-band channels increases (dashed blue curve in the C-band). However, there is no significant improvement in GSNR and TC in the C-band and LC-band scenarios when comparing FRP to FLP. Furthermore, as illustrated in Fig. \ref{fig:GSNR_OSNR_Freq_ALL_70km}(b), an approximate flat OSNR per band can be observed, which is advantageous for operation, administration, and management tasks.
The FRP in the LCS-band scenario shows promising results, with a maximum GSNR gain of about 2.5 dB and an OSNR gain of about 6 dB in the S-band. However, while the GSNR of the L-band channels decreases, this reduction, as demonstrated in the network-wide study, does not necessitate a change in the modulation format of the LPs. Furthermore, we observe an increase in GSNR for the LC-band scenario. This implies that in a pay-as-you-grow migration from the LC band to the LCS band, there is no bit rate penalty for most of the C-band channels. Additionally, we see an approximate OSNR profile for each band. In Figs. \ref{fig:GSNR_OSNR_Freq_ALL_70km}(a) and (b), there are six curves corresponding to each band and two HPO algorithms. Boxplots representing these six curves for each band are illustrated in Figs. \ref{fig:GSNR_OSNR_Freq_ALL_70km}(c) and (d) to demonstrate their sensitivity. The mean ($\times$), median (-), maximum, minimum, and distribution of the channels' GSNR and OSNR are depicted. For instance, as depicted in Figs. \ref{fig:GSNR_OSNR_Freq_ALL_70km}(c), (e), and (g), FRP not only flattens the GSNR but also significantly improves the GSNR of the S-band channels by about 2.5 dB.  
Now, let us evaluate FRP versus FLP for various span lengths. In a real-world network, span lengths vary, leading to differences in physical layer parameters such as amplifier noise figure and loss coefficient, among others, which may change due to aging. This necessitates rapid calculation of the optimum pre-tilt launch power. Additionally, as depicted in Fig. \ref{fig:POWER_FREQ_50100} (a), the tilt is not fixed in each band. FRP presents a promising approach to determine the tilt and offset of each band. In the FLP scenario, illustrated in Fig. \ref{fig:POWER_FREQ_50100} (b), the tilt power is adjusted at the receiver, whereas in the FRP scenario, the tilt launch power is adjusted (Fig. \ref{fig:POWER_FREQ_50100} (a)). Once again, we observe the seesaw effect in Fig. \ref{fig:POWER_FREQ_50100}, where achieving a flat power at the receive side requires increasing the launch power of higher frequency channels. However, considering the dynamic range of doped fiber amplifiers (DFAs) and managing NLI, the maximum launch power has been restricted to 6 dBm \cite{MultiBand_Amplifier_JLT}. Hence, we observe fewer changes in S-band channels, particularly for distances of 80-100 km. As illustrated in Fig. \ref{fig:POWER_FREQ_50100}(a), the slope of the tilt in the L-band differs significantly from that of the C- and S-bands. Consequently, the tilt-offset approach proposed in \cite{Correia_PowerControl, Luo_22_optic_express} becomes highly time-consuming when determining the optimal launch power for live networks, especially when physical parameters of the span are altered. As anticipated, the optimal power also increases with an increase in the span length. The optimal launch powers are within the ranges of [-0.8, 0.9] for FLP and [-6, 6] for FRP. Box plots in Fig. \ref{fig:POWER_FREQ_50100} (c) and (d) present the statistical data of channels' GSNR and OSNR for both FLP and FRP, respectively, for each span length. As demonstrated, the gain of FRP over FLP increases with the span length, resulting in up to 6 dB maximum improvements in GSNR and OSNR over a 100 km span. On average, the improvement in GSNR and OSNR is approximately 0.7 dB for a 100 km span. Notably, the QoT gain after 70 km does not change significantly due to the power threshold that has been established. 
%height = 1\linewidth, 
Subsequently, we present two significant metrics: standard deviation (STD) and the disparity between the maximum and minimum channels' GSNR/OSNR, denoted as Max-Min(GSNR/OSNR). Lower values of STD and Max-Min(GSNR/OSNR) indicate a flatter GSNR/OSNR profile across channels. These flatness indicators are depicted in Fig. \ref{fig:GSNR_OSNR_STD_MAXMIN_50100}, where the dashed curves associated with the FRP approach demonstrate lower STD and Max-Min values. Therefore, FRP enhances the average QoT of a span and provides a more uniform GSNR/OSNR profile. Furthermore, Table \ref{tab:TOCT_spans50_100} presents the TC of spans ranging from 50 to 100 km. It is evident that FRP yields more significant gains for longer span lengths, increasing from 0.4 Tbps for 50 km to approximately 10 Tbps for 100 km.
% \begin{figure}[!t]
% \centering
% \includegraphics[width=1\linewidth]{Figures/POWER_FREQ_50100_2}
% \caption{Simulation results for the network.}
% \label{fig:POWER_FREQ_50100_2}
% \end{figure}
\begin{table}[!t]
\centering
\caption{Total capacity (TC) [\lowercase{Tbps}] for flat launch power (FLP) and FRP (flat receive power) power optimization in terms of span lengths. }
\label{tab:TOCT_spans50_100}
\begin{tabular}{|c|c|c|c|c|c|c|}
\hline
Span Length [km] & 50    & 60    & 70    & 80    & 90    & 100   \\ \hline
FLP              & 445.0 & 421.8 & 398.7 & 375.5 & 352.0 & 328.2 \\ \hline
FRP              & 445.3 & 425.1 & 405.0 & 384.6 & 362.1 & 337.8 \\ \hline
Diff [Tbps]      & 0.4   & 3.3   & 6.3   & 9.0   & 10.0  & 9.7   \\ \hline
\end{tabular}
\end{table}

\subsection{Network-wide Study}\label{Network_wideStudy}

\begin{figure*}[!t]
\centering
\includegraphics[width=1\linewidth]{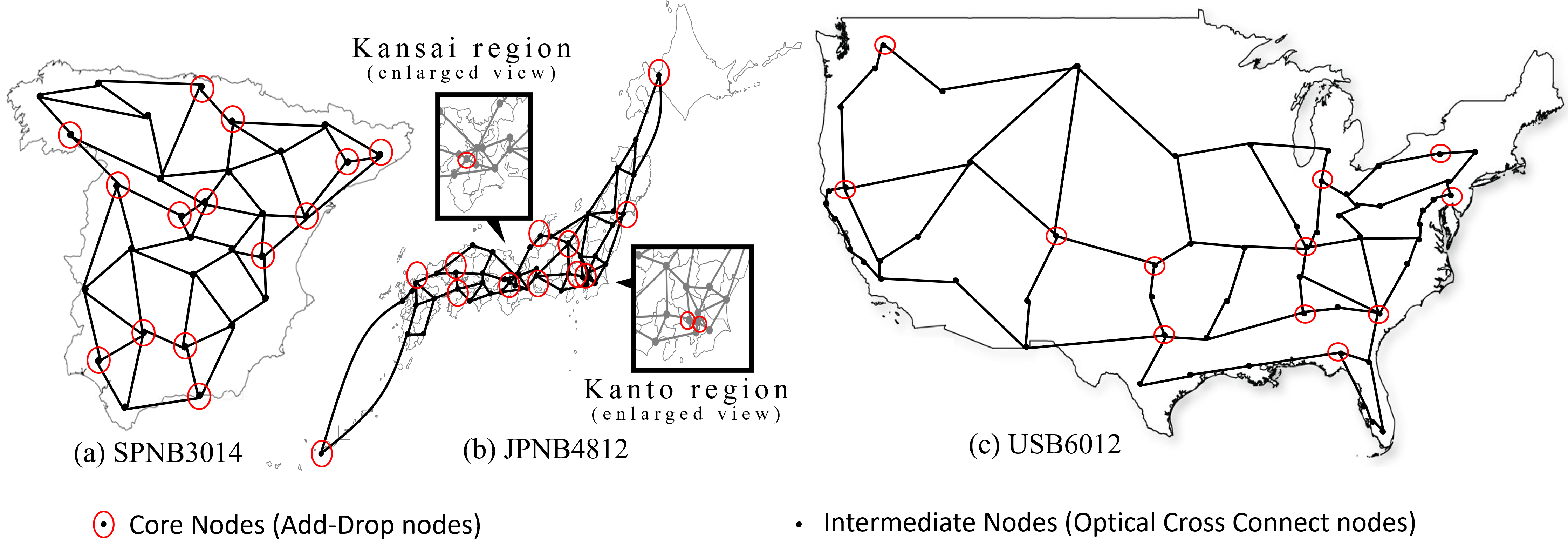}
\caption{Network topology of (a) Spanish backbone (SPNB3014), (b) Japanese backbone (JPN4812), and (c) United State backbone (USB6012).}
\label{fig:Topologies}
\end{figure*}

\begin{figure}[!t]
\centering
\includegraphics[width=1\linewidth]{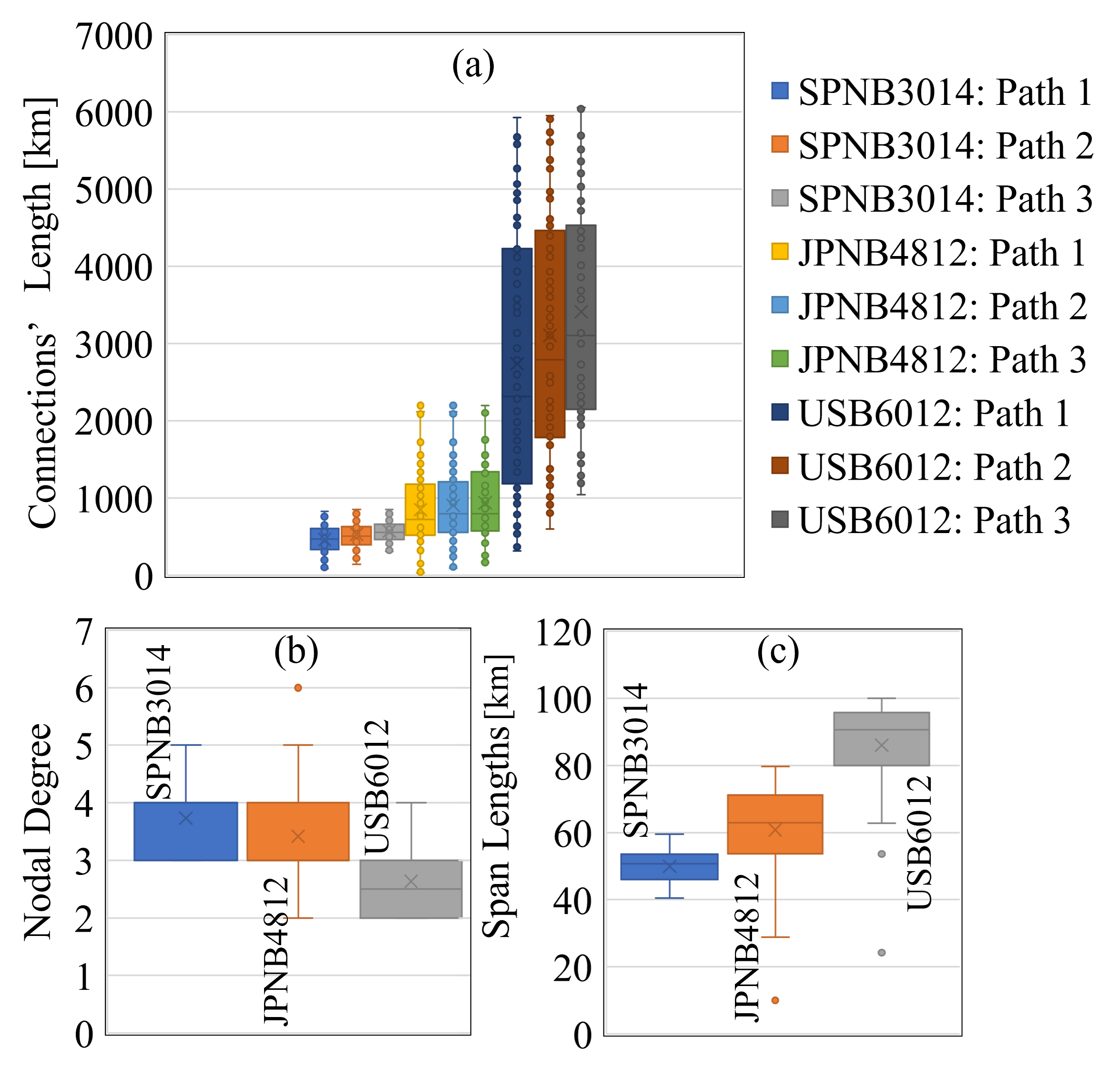}
\caption{(a) Connections' length, (b) Nodal degree, and (c) span length of SPNB3012, JPNB4812, and USB6012.}
\label{fig:PathDistances_AllNetworks}
\end{figure}

This section extensively elaborates on comparing the performance between FLP and FRP at the network-wide level. Metrics such as bandwidth blocking probability (BBP), modulation format cardinality usage, average GSNR, and the number of LCIs have been selected to demonstrate the effectiveness of each HPO approach. To achieve this, we have examined three networks based on the length of the LPs. These networks include the Spanish backbone (SPNB) with 30 nodes and 56 links, the Japanese backbone (JPNB) with 48 nodes and 82 links, and the United States backbone (USB) with 60 nodes and 79 links, as depicted in Fig. \ref{fig:Topologies}. To represent a realistic backbone network planning scenario, we designated specific nodes as the core of the backbone networks. In contrast, the remaining nodes are intermediate nodes equipped solely with ROADM functionalities without add/drop capabilities. Essentially, these intermediate nodes serve as metro-core nodes, where their traffic is aggregated before reaching the core nodes. This assumption allows us to create a more realistic scenario that accounts for filtering penalties at the intermediate nodes. Accordingly, we refer to the topologies as SPNB3014, JPNB4812, and USB6012. In SPNB, 14 out of 30 serve as core nodes, in JPNB, 12 out of 48, and in USB, 12 out of 60. The total number of bidirectional connections, comprising all source-destination pairs, in each network is calculated as $N(N-1)/2$. The selection of a source-destination pair is based on a biased probability distribution function determined by geographical population and degree number \cite{Correia_PowerControl}. The box plot illustrating the lengths of the three shortest paths depicted in Fig. \ref{fig:PathDistances_AllNetworks} (a). 
In Fig. \ref{fig:PathDistances_AllNetworks} (a), it is evident that the maximum lengths of connections for SPNB3014 are under 1000 km, for JPNB4812 they are under 2350 km, and for USB6012 they are under 6150 km. The maximum span lengths for SPNB3014, JPNB4812, and USB6012 are 60 km, 80 km, and 100 km, respectively. Nodal degree and span length statistics are depicted in box plots in Fig. \ref{fig:PathDistances_AllNetworks} (b) and (c). This figure illustrates the importance of considering the ROADM filtering effect and the heterogeneous span lengths in planning, which has been overlooked in the distance adaptive approach. Further details regarding the networks under study in this paper can be found in \cite{Telefonica}, \cite{JPNNetwork}, and \cite{simmons2014optical} for SPNB3014, JPNB4812, and USB6012, respectively. 

\subsubsection{Channel-Connection Capacity Profile}\label{CCR}

\begin{figure}[!t]
\centering
\includegraphics[width=0.8\linewidth]{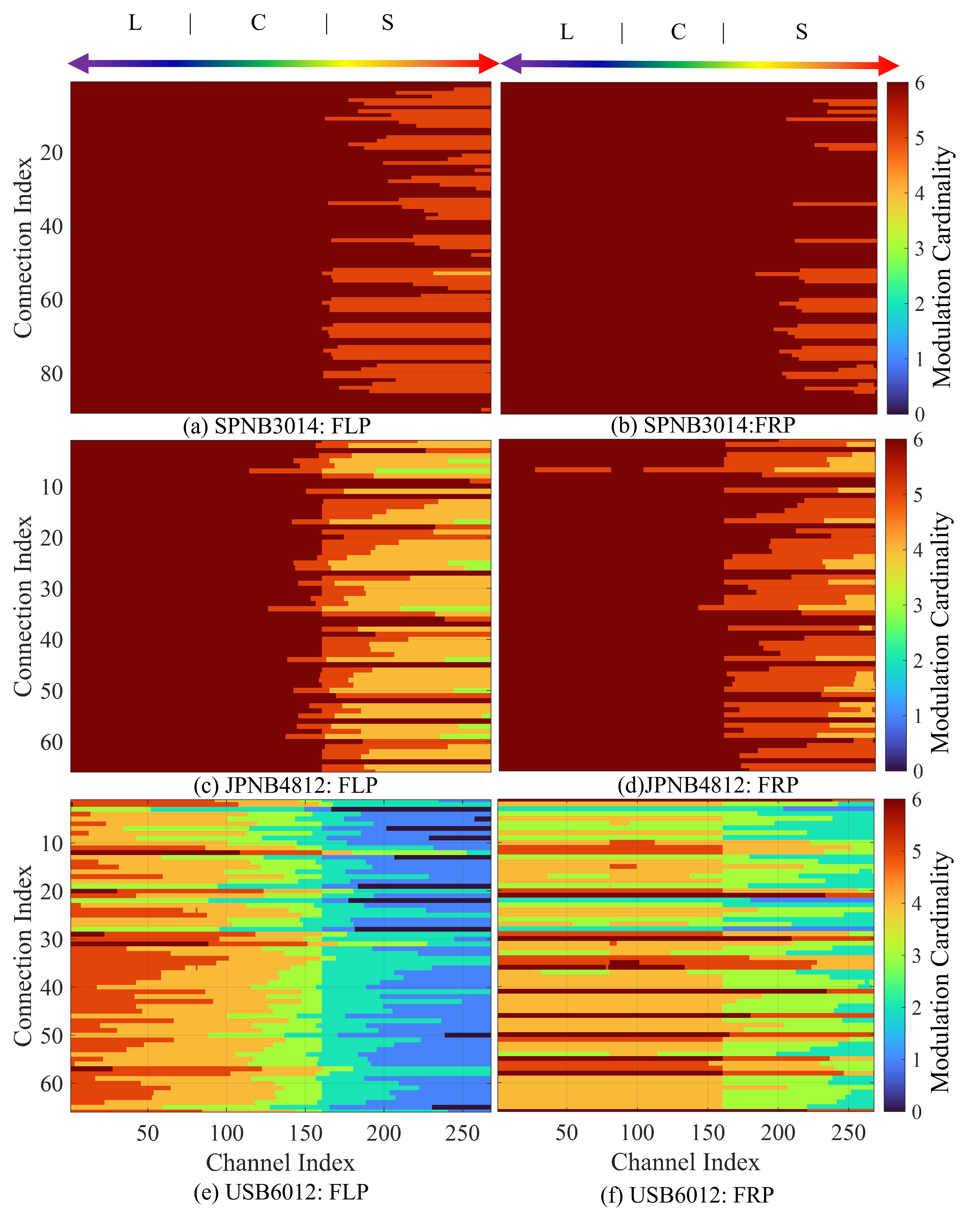}
\caption{Channel-connection capacity profile for the first candidate path in (a) SPNB3014-FLP, (b)SPNB3014-FRP, (c)JPN4812-FLP, (d)JPN4812-FRP, (e)USB6012-FLP, (f)USB6012-FRP. $^*$ to save space, we skip to show the channel-connection capacity profiles of the second and third shortest paths.}
\label{fig:Modulation_perChannel_AllNetworks}
\end{figure}

The wavelength-dependent capacity profile of all connections for the first candidate shortest path is illustrated in Fig. \ref{fig:Modulation_perChannel_AllNetworks} (a)-(f) for both FLP and FRP HPOs across three networks. It is observed that higher modulation cardinality usage occurs in FRP mode for both C- and S-band channels. As discussed later, modulation cardinality ranges from 1 to 6, corresponding to bit rates of 100G to 600G for LCIs, with these values determined based on the GSNR of connections in each band. Fig. \ref{fig:Modulation_perChannel_AllNetworks} demonstrates that the bit rate of an LCI for a connection depends entirely on the channel frequency. Additionally, it can be concluded that in the SPNB3014 and JPN4812 networks, most of the channel-connection resources (CCRs) in the L-band operate at the highest modulation cardinality, maintaining the same bit rate in both FLP and FRP modes. While FRP may degrade the L-band channels, the connection distances and GSNR thresholds remain within safe limits, with no observed bit rate degradation. However, in some instances, such as connection 8 in JPNB4812 and very long-haul connections in USB6012, the bit rate of L-band CCRs in FLP mode may be higher than in FRP mode. Nonetheless, a different behavior is observed concerning C- and especially S-band CCRs. For instance, as illustrated in Fig. \ref{fig:Modulation_perChannel_AllNetworks}, there are certain CCRs in the S-band for USB6012 in FLP mode that cannot be utilized due to their GSNR being lower than the minimum GSNR required for the lowest modulation cardinality (as depicted by black markers in Fig. \ref{fig:Modulation_perChannel_AllNetworks} (e)). However, when transitioning to FRP mode, these CCRs survive, albeit with a decrease in bit rate in the L-band (as shown in Fig. \ref{fig:Modulation_perChannel_AllNetworks} (f)). Consequently, the increase in GSNR in the S-band is offset by the decrease in GSNR in the L-band. Notably, the number of CCRs in the S-band exceeds those in the L-band. Therefore, studying FLP and FRP at a network-wide level is imperative to quantify the benefits FRP offers to telecommunication operators (Telcos).

% To examine the fluctuations in modulation cardinality efficiency, we introduce the channel-connection efficiency (CCE) rate as $\sum_c\sum_i R^{l,i}$, where $R^{l,i}$ denotes the bit rate of connection $c$ when it utilizes channel $i$ across all its links. As illustrated in Fig. \ref{fig:CCE_delta}, the CCE for SPNB3014 and JPNB4812 exhibits an increase from path 1 to path 3. This implies that not only do the candidate shortest paths have relatively short distances, but also modulation cardinality changes occur at higher levels. However, in the case of USB6012, we observe the opposite trend due to the longer distances among the first (second) shortest paths and the third ones.

% \begin{figure}[!t]
% \centering
% \includegraphics[width=1\linewidth]{Figures/CCE_delta
% }
% \caption{The differential channel-connection efficiency (CCE) of the FRP and FLP method.}
% \label{fig:CCE_delta}
% \end{figure}
\subsubsection{The Networks' Optimal Launch Power profiles per Span}\label{HPO_SpanLevel}
\begin{figure}[!t]
\centering
\includegraphics[width=1\linewidth]{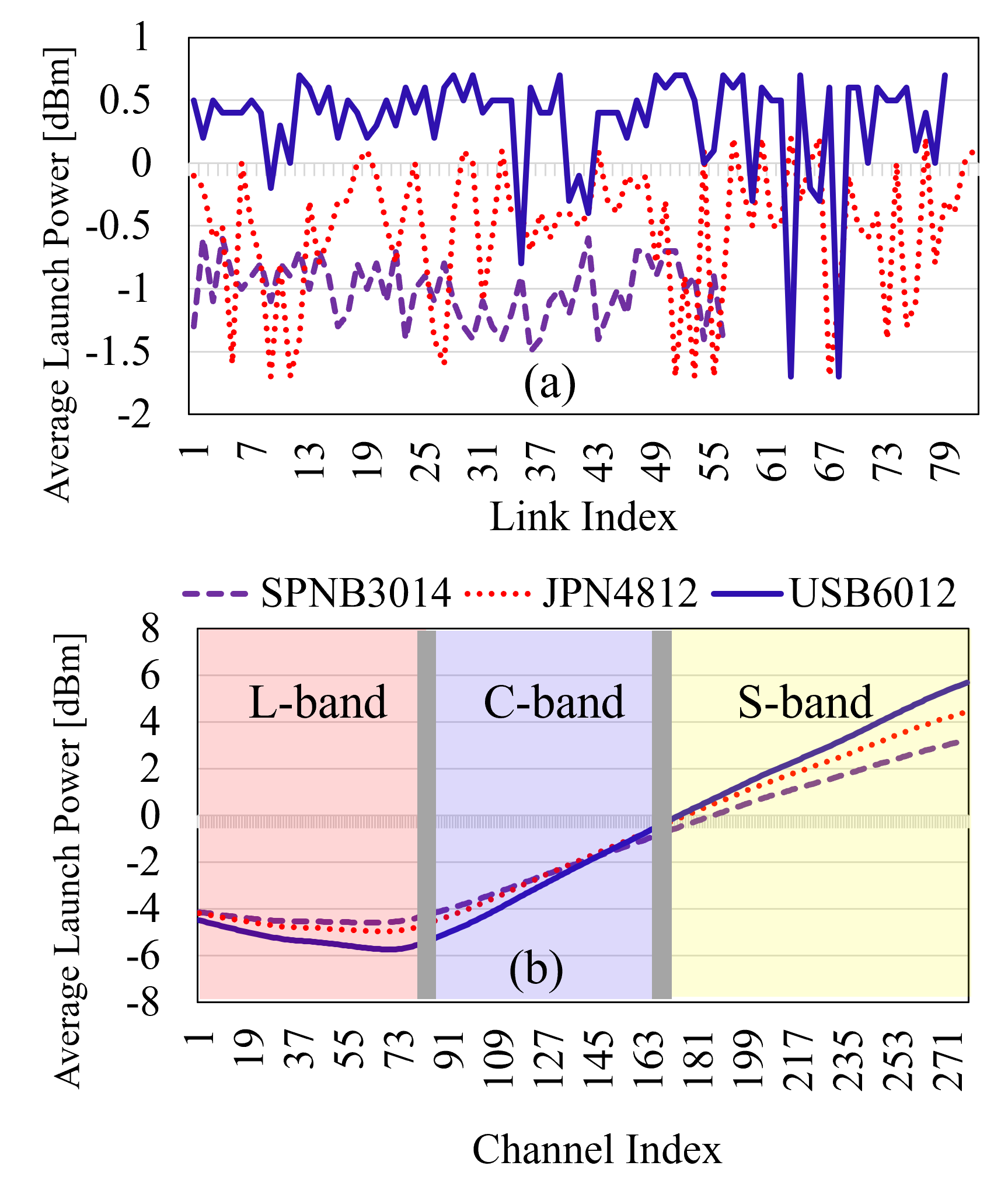}
\caption{The average launch power (a) per link is optimized with flat launch power (FLP), and (b) per links and channels is optimized with flat received power (FRP) in SPNB3014, JPN4812, and USB6012.}
\label{fig:Power_Networks}
\end{figure}

The launch power profile of the FLP and FRP modes are depicted in Fig. \ref{fig:Power_Networks} (a) and (b) respectively. As anticipated, both the average and range of launch power tend to increase with network size. In the FLP mode (Fig. \ref{fig:Power_Networks} (a)), the SPNB3014 exhibits a launch power range between -1.5 dBm and -0.5 dBm, the JPN4812 between -1.7 dBm and 0.1 dBm, and the USB6012 between -1.7 dBm and 0.7 dBm. Given that the span length is consistent for each link and we have considered the average span length for each link, the optimal power for each link is reported. In contrast, in the FRP mode, where pre-tilted launch power is employed, the average launch power profiles for all links per network are presented in Fig. \ref{fig:Power_Networks} (b). The behavior of the launch power profile differs in this case; with increasing network size, the launch power range expands in the S-band but contracts in the C- and L-bands, exhibiting a seesaw effect. 
\subsubsection{GSNR and Modulation Cardinality Usage}\label{GSNR_MFL}
\begin{figure*}[!t]
\centering
\includegraphics[width=1\linewidth]{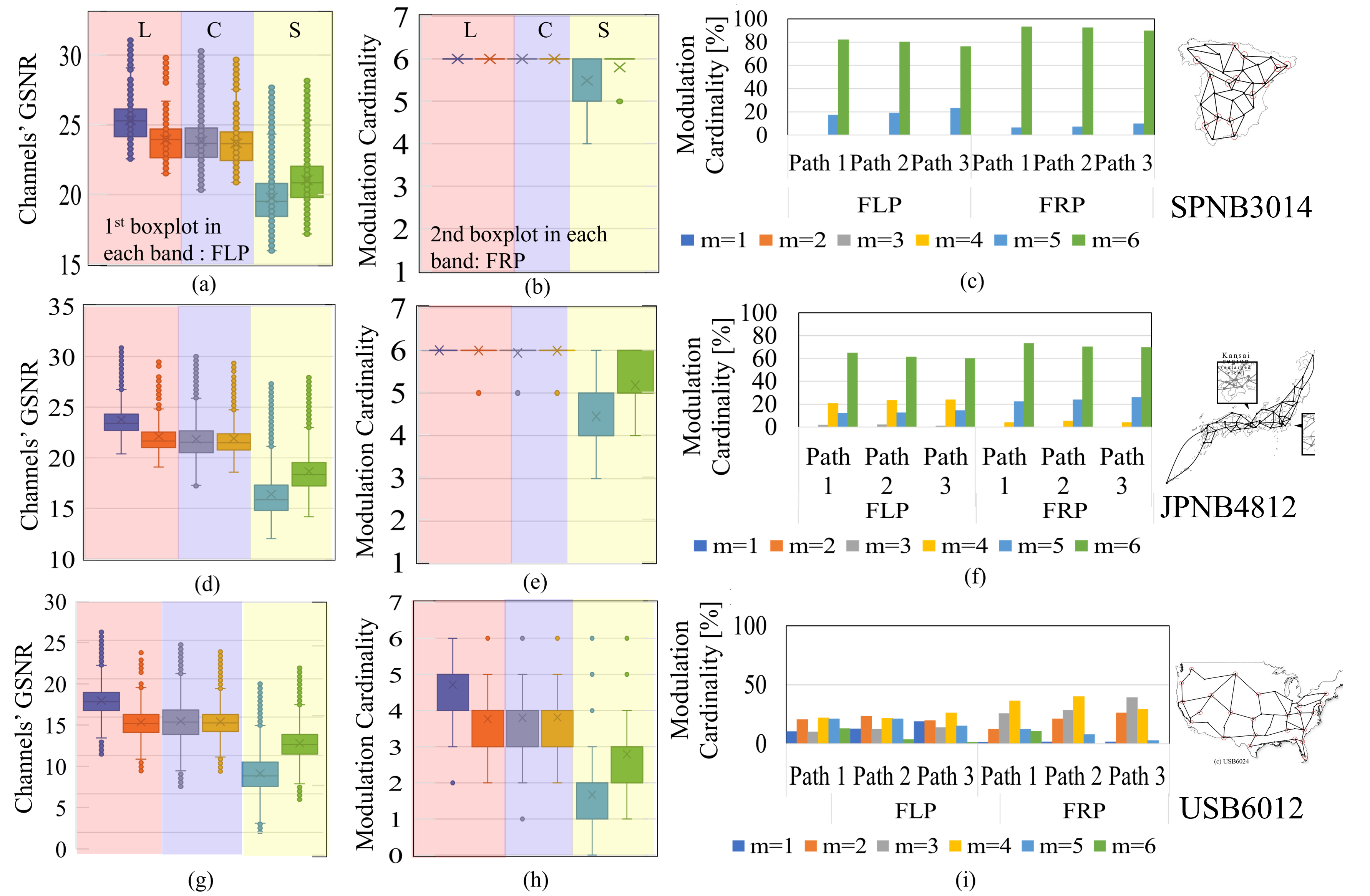
}
\caption{GSNR and modulation cardinality box plots, and modulation cardinality percentage in each candidate path with FLP and FRP for (a)-(c) SPNB3014, (d)-(f) JPNB4812, and (g)-(i) USB6012.}
\label{fig:GSNR_Modulation_allNetwork}
\end{figure*}

To comprehensively understand FLP and FRP performance in MB-EONs, it is essential to scrutinize GSNR and modulation cardinality characteristics. Figs \ref{fig:GSNR_Modulation_allNetwork} (a)-(i) depict GSNR traits for all CCRs, along with modulation cardinality usage and percentage of modulation cardinality usage per shortest path for SPNB3014 ((a)-(c)), JPNB4812 ((d)-(f)), and USB6012 ((g)-(i)). Notably, the first box plot in each band corresponds to FLP, while the second represents FRP. Although the average GSNR of L-band CCRs decreases in FRP compared to FL, no modulation cardinality changes are observed in SPNB3014 across all shortest paths (1, 2, and 3). Conversely, JPNB4812 and USB6012 demonstrate distinct trends, showcasing, for instance, a 3 dB average GSNR degradation and 1.3 modulation cardinality changes on average for USB6012, and 1 dB and 1, respectively, for JPNB4812. 
An intriguing discovery is the comparable performance of FRP and FLP in terms of the average GSNR and modulation cardinality in the C-band across all candidate paths. FRP exhibits promising performance in the S-band, enhancing both average GSNR and utilization of higher modulation cardinality. Consequently, CCR capacity increases significantly across all candidate shortest paths, with USB6012 particularly benefiting due to its numerous long-haul connections, such as those spanning the eastern and western coasts of the USA. Notably, FRP enables the survival of previously infeasible CCRs and eliminates CCRs with zero modulation cardinality. 
Lastly, the percentage of modulation cardinality usage per candidate path, depicted in the last column of Fig. \ref{fig:GSNR_Modulation_allNetwork}, underscores a consistent enhancement in CCRs capacity with significant modulation cardinality changes from lower to higher levels in FRP compared to FLP. 
\subsubsection{Bandwidth Blocking Probability Analysis}\label{BBPAnalysis}
The primary question that often arises is how FRP can potentially decrease the BBP in the network. Answering this question can provide Telcos with valuable insights to make informed decisions between FRP and FLP, potentially reducing the implementation costs of network infrastructure. However, it is worth noting that higher GSNR, especially in the S-band, is guaranteed with FRP, thereby ensuring higher modulation cardinality for CCRs. Based on previous results, we can speculate that FLP might not significantly underperform, depending on the network size. To explore this further, we adopt a semi-static approach, commonly used in backbone networks, where traffic remains in the network for an extended period once it arrives. Our analysis considered k=3 shortest candidate paths for each request. The CCRs were pre-calculated for FLP and FRP scenarios, with results obtained from 1000 random iterations. The requested bit rate ranged randomly from 100 Gbps to 600 Gbps, with Flexponder grooming taken into account based on Fig. \ref{fig:Grooming_Multiband}. However, it is noted that the authors in \cite{Correia_PowerControl} and \cite{JanaJOCN022QoT} did not consider Flexponder grooming, prompting us to remodel their algorithm accordingly. We implement a minimum LCI deployment algorithm, implying that if sufficient resources are available among the already installed LCIs for a connection, deploying new LCIs is unnecessary. Notably, to minimize jitter, we assume that the flow of each request wholly traverses a single path. Two path selection algorithms are examined: one focuses on maximizing the minimum GSNR of the LCIs, named MaxMinGF, while the other aims at minimizing the maximum LCIs' frequency, named MinMAXF. Additionally, three modulation cardinality selection algorithms are explored: channel-based GSNR (CBG), worst case in all bands (WAB), and worst case per band (WPB). In CBG, we evaluate the modulation cardinality for each channel based on channel-connection resources. In WAB \cite{SamboJLT2020,commletter_distanc,MehrabiJLT2021}, a distance adaptive approach is adopted, considering the minimum GSNR across all bands. We employ a similar distance adaptive approach in WPB but focus on the minimum GSNR within each band for modulation cardinality selection \cite{commletter_distanc,commletter_distanc,MehrabiJLT2021}. 

The results from the BBP are demonstrated in Fig. \ref{fig:BBPLOGN_GON}(a)-(c). Since a 1\% BBP is deemed acceptable for Telcos, these algorithms are evaluated against BBP  1\%. As expected, FRP demonstrates superior performance compared to FLP across all networks. The network throughput gain of FRP over FLP varies depending on the network size, with an approximate gain of 5\% observed in all networks under study. Interestingly, despite the increase in average GSNR and CCRs with FRP at both the span and CCR levels, there is no significant enhancement in network throughput in backbone terrestrial networks. Moreover, FRP could be more expensive than the FLP scenario due to DGEs and software issues. Hence, this aspect warrants further exploration in future studies, which may be of interest from Telcos' perspective.
In conclusion, a maximum 10 Tbps improvement in TC per span may not significantly enhance the total network throughput. Regarding the path selection algorithm, MinMaxF displayed a more significant performance than MaxMinGSNRF, with a notable decrease in spectral fragmentation observed. As anticipated, CBG, based on synergizing the HPO and wavelength-based QoT-aware cross-layer design, demonstrates a network throughput gain of up to 75\% at a 1\% BBP compared to WAB and WAP.
\subsubsection{Local Optimization versus Global Optimization}\label{LOGONvsGON}
\begin{figure*}[!t]
\centering
\includegraphics[width=1\linewidth]{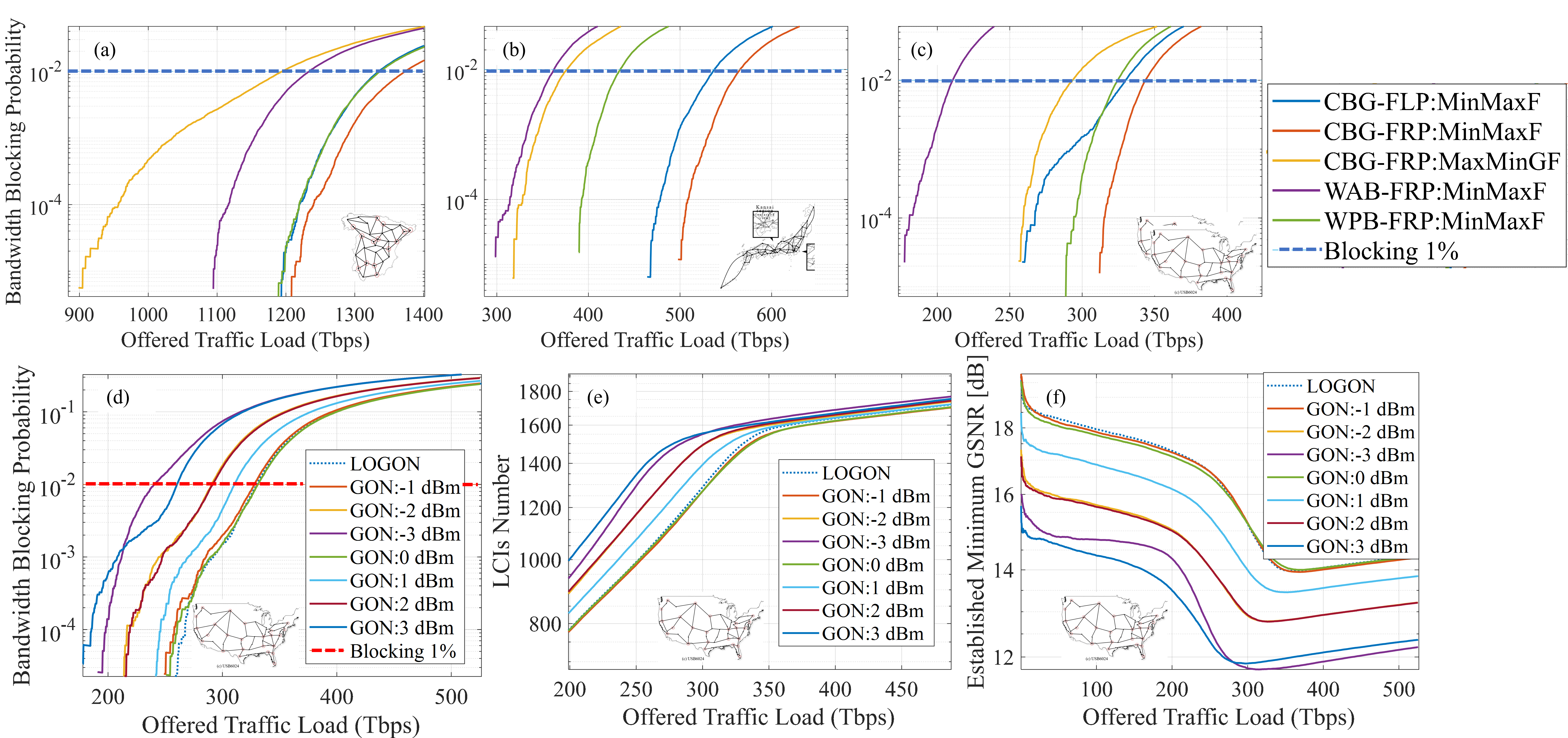}
\caption{The bandwidth blocking probability (BBP) v.s offered traffic load (OTL) for different path selection and power optimization methods (a) SPNB3014, (b) JPN4812, and (c) USB6012. The BBP, line card interfaces number, and GSNR v.s OTL for LOGN and GON in (d), (e), and (f), respectively. }
\label{fig:BBPLOGN_GON}
\end{figure*}

In the latest analysis, we compared the traditional LOGON power optimization and the proposed global power optimization at the network-wide level (GON). While LOGON (optimum power per span) is designed for specific cases in C-band optical networks, its performance, alongside GON, has not been evaluated at the network-wide level in terms of BBP. In the GON approach, we aim for uniform optimal power settings across all spans to enhance network throughput at a given BBP (e.g., 1\%). For this study, we employ the FLP algorithm in USB6012 without loss of generality. As illustrated in Fig. \ref{fig:BBPLOGN_GON}(d)-(f), GON with launch powers of 0 and -1 dBm demonstrates comparable performance to FLP LOGON. Given that implementing GON in hardware and software is simpler and potentially more cost-effective than LOGON, GON is recommended. The BBP, LCI usage, and average GSNR of established LPs versus offered traffic load are depicted in Fig. \ref{fig:BBPLOGN_GON} (d)-(f), respectively. 

% \begin{figure*}[!t]
% \centering
% \includegraphics[width=0.7\linewidth]{Figures/Blockin_LCI_GSNR_FLP_vs_Powers}
% \caption{The bandwidth blocking probability v.s offered traffic load for different path selection and power optimization methods (a) SPNB3014, (b) JPN4812, and (c) USB6012.}
% \label{fig:Blockin_LCI_GSNR_FLP_vs_Powers}
% \end{figure*}

\section{Conclusion}\label{sec_VI_conclusion}
The paper explored various power optimization strategies for multi-band elastic optical networks (MB-EONs), focusing on fixed-grid flexible bit rate LCIs capable of adapting to different modulation cardinalities based on the GSNR of the channel. It compares two hyper-accelerated power optimization (HPO) algorithms based on the local optimization-global optimization (LOGO) concept: flat Receive Power (FLR) and flat Launch Power (FLP). Through exhaustive simulations conducted at both the span and network-wide levels, the study reveals that while FRP introduces greater total capacity in terms of span-level and channel-connection resources, its performance does not significantly outperform FLP across networks of varying sizes. The network gain achieved by FRP versus FLP is approximately 5\%. Additionally, a wavelength-dependent QoT-aware modulation format cardinality assignment model, termed channel-based GSNR, demonstrates up to a 75\% higher network throughput compared to previously proposed worst-case scenarios. Lastly, the study indicates that employing identical power settings for all spans named global optimization of the network yields similar performance in terms of maximum network throughput at a fixed BBP in the network-wide analysis. This paper serves as a benchmark and provides valuable insights for telecommunication operators considering power optimization strategies when transitioning from C-band to beyond-C-band networks.
% \section*{Acknowledgments} 
% Farhad Arpanaei acknowledges support from the CONEX-Plus program funded by Universidad Carlos III de Madrid and the European Union's Horizon 2020 research and innovation program under the Marie Sklodowska-Curie grant agreement No. 801538. The authors from UC3M, Telefonica, and Infinera would like to acknowledge the support of the EU-funded ALLEGRO project (grant No.101016663). The authors from UC3M would
% like to acknowledge the support of the Spanish-funded Fun4date-Redes project (grant No.PID2022-136684OB-C21).
\bibliographystyle{IEEEtran}
\bibliography{IEEEabrv,REF}
\end{document}